\documentclass[aps,prd,superscriptaddress]{revtex4-2}
\usepackage{amsmath,amssymb,amsthm,bm,graphicx,subfigure,xcolor}
\usepackage[colorlinks=true,linkcolor=blue,citecolor=blue,urlcolor=black]{hyperref}

\begin{document}
	
\title{Dynamics of holographic steady flows near a first-order phase transition}

\author{Qian Chen}
\email{chenqian192@mails.ucas.ac.cn}
\affiliation{School of Physical Sciences, University of Chinese Academy of Sciences, Beijing 100049, China}
\affiliation{School of Fundamental Physics and Mathematical Sciences, Hangzhou Institute for Advanced Study, University of Chinese Academy of Sciences, Hangzhou, Zhejiang, 310024, China}
\affiliation{Beijing Institute of Mathematical Sciences and Applications, Beijing, 101408, China}

\author{Yuxuan Liu}
\email{liuyuxuan93@csu.edu.cn}
\affiliation{Institute of Quantum Physics, School of Physics, Central South University, Changsha 418003, China}

\author{Yu Tian}
\email{ytian@ucas.ac.cn}
\affiliation{School of Physical Sciences, University of Chinese Academy of Sciences, Beijing 100049, China}
\affiliation{Institute of Theoretical Physics, Chinese Academy of Sciences, Beijing 100190, China}

\author{Xiaoning Wu}
\email{wuxn@amss.ac.cn}
\affiliation{Institute of Mathematics, Chinese Academy of Sciences, Beijing 100190, China}

\author{Hongbao Zhang}
\email{hongbaozhang@bnu.edu.cn}
\affiliation{School of Physics and Astronomy, Beijing Normal University, Beijing 100875, China}
\affiliation{Key Laboratory of Multiscale Spin Physics, Ministry of Education, Beijing Normal University, Beijing 100875, China}

\begin{abstract}
We investigate the physical properties of steady flows in a holographic first-order phase transition model, extending from the thermodynamics at equilibrium to the real-time dynamics far from equilibrium.
Through spinodal decomposition or condensation nuclei, the phase-separated state with non-zero momentum can be achieved.
In this scenario, we observe a gap between coexisting phases, arising not only from the variations in energy density, but also from the distinctions in momentum density or longitudinal pressure. These disparities are characterized by flow velocity and latent heat.
Furthermore, by introducing an inhomogeneous scalar external source to simulate a fixed obstacle, we reveal the dynamical response of momentum loss in the moving system.
Notably, starting from an initial phase-separated state with uniform flow velocity, and subsequently interacting it with an obstacle, we find that the moving high-energy phase exhibits four characteristic dynamical behaviors---rebounding, pinning, passing, and splitting. These behaviors depend on the velocity of the phase and the strength of the obstacle.
\end{abstract}


\maketitle

\section{Introduction}\label{sec:I}
The dynamics of strongly coupled quantum fluids is of great research significance in particle physics experiments and astronomical observations, such as the transport properties of the quark gluon-plasma (QGP) \cite{Schafer:2009dj} and the merger process of binary neutron stars \cite{Alford:2017rxf}.
The challenge is to construct a theoretical framework to describe the real-time evolution of strongly interacting matter.
Fortunately, by dualizing a quantum many-body system to a classical gravitational entity with an extra dimension, holography \cite{Maldacena:1997re,Gubser:1998bc,Witten:1998qj,Witten:1998zw} converts such a problem into solving a system of time-dependent gravitational equations, which is always amenable to the numerical relativity \cite{Chesler:2013lia}.

A breakthrough related to holographic fluids is the description of dynamics near a thermal first-order phase transition.
Initially, in order to mimic the equation of state of quantum chromodynamics (QCD), the holographic first-order phase transition model was constructed bottom-up \cite{Gubser:2008yx,Gubser:2008ny}, involving a real self-interacting scalar field with an external source.
Subsequently, the dynamical stability of the thermal phases in equilibrium was revealed through linear perturbation theory \cite{Janik:2015iry,Janik:2016btb}.
The results indicate the existence of a branch of hydrodynamic modes that dominates the linear stability of the system.
For the spinodal region with negative specific heat, the hydrodynamic modes become dynamically unstable in the long wavelength region, similar to Gregory-Laflamme instability \cite{Gregory:1993vy,Gregory:1994bj}.
Going a step further, the real-time dynamics of the spinodal instability was numerically simulated at the nonlinear level \cite{Attems:2017ezz,Janik:2017ykj,Bellantuono:2019wbn,Attems:2019yqn,Attems:2020qkg,Caddeo:2024lfk}, revealing the second-order hydrodynamic description of the dynamical process and the phase-separated configuration of the final state.
On the other hand, for the thermodynamically metastable region adjacent to the spinodal region, the occurrence of a dynamical transition requires the system to cross an unstable critical state \cite{Bea:2020ees,Bea:2022mfb,Chen:2022cwi}, in which a critical phenomenon analogous to that in gravitational collapse \cite{Choptuik:1996yg,Liebling:1996dx,Bizon:1998kq,Gundlach:2007gc} occurs, indicating a nonlinear instability.
Furthermore, through holographic quench, the nonlinear instability of the phase-separated state was also revealed \cite{Chen:2022tfy}, achieving a bidirectional dynamical transition between homogeneous and inhomogeneous configurations.
At this point, the theoretical framework on the dynamical stability of the holographic first-order phase transition system has been roughly completed.

Beyond stability analysis, the dynamic properties of holographic fluids near phase transitions and their applications have also been extensively investigated.
An interesting research topic is the simulation of cosmological phase transitions \cite{Kibble:1980mv,Bigazzi:2020phm,Hindmarsh:2020hop}, with the resulting gravitational wave spectrum potentially observable to detectors \cite{Bigazzi:2020avc,Ares:2020lbt,Ares:2021nap}.
For the case of cosmological phase transitions proceed via bubble nucleation, the dynamic description of the domain wall connecting the nucleated bubble of a stable phase and the overcooled medium was successfully revealed \cite{Bea:2021zsu,Bigazzi:2021ucw,Janik:2021jbq,Janik:2022wsx}, in which the velocity of the domain wall, a crucial parameter on which the gravitational wave spectrum depends, is quantitatively characterized.
On the other hand, if the Universe supercools along the metastable branch and enters the spinodal region, there will be a cosmological phase transition induced by the exponential growth of unstable modes, resulting in a qualitatively different gravitational wave spectrum \cite{Bea:2021zol}.
Another important application lies in the QGP produced in relativistic heavy ion collisions \cite{Casalderrey-Solana:2011dxg}, which are viewed as collisions of gravitational shock waves in the dual description \cite{Chesler:2010bi,Casalderrey-Solana:2013aba,Casalderrey-Solana:2013sxa,Chesler:2015wra,Chesler:2015bba,Chesler:2016ceu}.
If QCD possesses a critical point, then the far-from-equilibrium dynamics of QGP across a phase transition can be effectively captured by a holographic model with a phase transition \cite{Attems:2016tby,Attems:2018gou,Bea:2021ieq}.
Other related content includes holographic turbulence \cite{Adams:2012pj,Adams:2013vsa}, holographic superfluid \cite{Herzog:2008he,Bhaseen:2012gg,Zhao:2023ffs}, etc.

Inspired by the fact that fluids in reality generally have a flow velocity, we will study the real-time dynamics of the holographic steady flow, which is dual to the boosted black brane geometry on the gravity side \cite{Bhattacharyya:2007vjd,Ecker:2021ukv},  near a thermal first-order phase transition. 
The plan of the paper is as follows.
In the following section \ref{sec:Hm}, we will introduce a holographic first-order phase transition model through the thermodynamic properties of the equilibrium system. 
In section \ref{sec:Ps}, we will study the dynamical transitions of holographic steady flows through fully nonlinear numerical simulations, revealing spontaneous and critical dynamical processes, which correspond to unstable and metastable initial states, respectively.
Similarly, the final state is a two-phase coexisting geometric configuration.
The difference is that the high-energy phase moves in the low-energy bath at a uniform velocity.

A natural question is whether there is a damping mechanism that slows down the moving domain.
We will provide such a mechanism in section \ref{sec:Is} by introducing an inhomogeneous external source to the scalar field.
From the Ward-Takahashi identity, such a local inhomogeneous scalar source can effectively induce changes in the momentum of the system, thereby damping the motion of a steady flow or a moving domain, which is verified by the dynamical results.
In particular, we reveal four characteristic dynamical behaviors of the moving domain under interaction with the local potential barrier: rebounding, pinning, passing through, and splitting, depending on the flow velocity of the moving domain and the strength of the potential barrier.
Finally, we end this paper with a conclusion and outlook in section \ref{sec:C}.


\section{Holographic model}\label{sec:Hm}
In this section, we will introduce a holographic model with a thermal first-order phase transition from equations of motion, to Ward-Takahashi identity, to thermodynamic phase diagrams involving static states and steady flows.

\subsection{Gravity setup}
{We consider a four-dimensional holographic model involving gravity and a scalar field, described by the following renormalized action \cite{Gibbons:1976ue,Bianchi:2001kw,Elvang:2016tzz}}
\begin{equation}
	2\kappa^{2}_{4}S_{\text{ren}}=\int_{M}dx^{4}\sqrt{-g}\mathcal{L}+2\int_{\partial M}dx^{3}\sqrt{-\gamma}K[\gamma]-\int_{\partial M}dx^{3}\sqrt{-\gamma}\left(R[\gamma]+4+\frac{1}{2}\phi^{2}\right),\label{eq:2.1}
\end{equation}
where $R[\gamma]$ is the Ricci scalar associated with the boundary induced metric $\gamma_{\mu\nu}$, $K[\gamma]$ is the trace of extrinsic curvature $K_{\mu\nu}=\gamma^{\sigma}_{\mu}\nabla_{\sigma}n_{\nu}$ with $n_{\nu}$ the outward normal vector field to the boundary, and bulk Lagrangian $\mathcal{L}$ describes an Einstein-scalar system
\begin{equation}
	\mathcal{L}=R-\frac{1}{2}\nabla_{\mu}\phi\nabla^{\mu}\phi-V(\phi),
\end{equation}
with a self-interaction potential
\begin{equation}
	V(\phi)=-6\text{cosh}\left(\frac{\phi}{\sqrt{3}}\right)-\frac{\phi^{4}}{5}.\label{eq:3}
\end{equation}
{According to the AdS/CFT correspondence, such a gravitational system is dual to a conformal field theory deformed with a scalar operator at the three-dimensional boundary.
The reason we shy away from working on a five-dimensional holographic model dual to a four-dimensional field theory is to avoid the appearance of holographic anomaly \cite{Bianchi:2001kw}, which would induce the field functions to exhibit a logarithmic asymptotic behavior at the infinite boundary, thus greatly reducing the accuracy and convergence of the numerical method.
On the other hand, the dynamics of the system depends on the form of the self-interaction of the scalar field \cite{Gubser:2008ny}.
Such self-interaction (\ref{eq:3}) will cause the phase diagram structure of the model to have a spinodal region, where the states suffer from a long-wavelength instability \cite{Janik:2015iry}, leading to the occurrence of first-order phase transition dynamics.
Such a setup can approximately capture the physical phenomena in the general case.
In order to better match QCD, a five-dimensional holographic model should be considered, in which the selection rules of the self-interaction potential are derived from the QCD data.
More generally, one can consider the general case of additional electromagnetic fields \cite{Cai:2022omk}.
These extensions bring certain challenges to numerical techniques.}

By performing variation on the above renormalized action (\ref{eq:2.1}), 
\begin{equation}
	2\kappa^{2}_{4}\delta S_{\text{ren}}=\int_{M} d^{4}x\sqrt{-g}\left(E^{g}_{\mu\nu}\delta g^{\mu\nu}+E^{\phi} \delta\phi\right)-\int_{\partial M}d^{3}x\sqrt{-\gamma_{0}} \left(\left\langle T_{ij}\right\rangle\delta\gamma_{0}^{ij}-2\left\langle O\right\rangle\delta\phi_{0}\right),\label{eq:2.4}
\end{equation}
where the subscripts ``0" denote the external sources of the corresponding fields on the boundary,
one can obtain the equations of motion of the gravitational field and the scalar field in the bulk 
\begin{subequations}
\begin{align}
	E^{g}_{\mu\nu}&= R_{\mu\nu}-\frac{1}{2}Rg_{\mu\nu}-\left[\frac{1}{2}\nabla_{\mu}\phi\nabla_{\nu}\phi-\left(\frac{1}{4}\left(\nabla\phi\right)^{2}+\frac{1}{2}V(\phi)\right)g_{\mu\nu}\right]=0,\\
	E^{\phi}&=\nabla^{\mu}\nabla_{\mu}\phi-\frac{dV(\phi)}{d\phi}=0,
\end{align}\label{eq:2.5}
\end{subequations}
and the corresponding operators on the boundary
\begin{subequations}
\begin{align}
	\left\langle T_{ij}\right\rangle&=\lim_{r\rightarrow \infty}r\left[R[\gamma]_{ij}-\frac{1}{2}R[\gamma]\gamma_{ij}-K_{ij}-\left(2-K+\frac{1}{4}\phi^{2}\right)\gamma_{ij}\right],\label{eq:2.6a}\\
	\left\langle O\right\rangle &=-\frac{1}{2}\lim_{r\rightarrow \infty}r^{2}\left(\phi+ n^{\mu}\nabla_{\mu}\phi\right).
\end{align}\label{eq:2.6}
\end{subequations}
For a geometry that satisfies the equations of motion (\ref{eq:2.5}), the invariance of the on-shell action (\ref{eq:2.4}) under the following diffeomorphism
\begin{equation}
	\delta \gamma^{ij}_{0}=\pounds_{\xi}\gamma^{ij}_{0},
	\quad \delta\phi_{0}=\pounds_{\xi}\phi_{0},
\end{equation}
where $\pounds_{\xi}$ is the Lie derivative with respect to an arbitrary vector field $\xi^{i}$ tangent to the boundary, implies the Ward-Takahashi identity for the energy-momentum tensor of the boundary theory
\begin{equation}
	\nabla^{j}\left\langle T_{ij}\right\rangle=\left\langle O\right\rangle \nabla_{i}\phi_{0},\label{eq:2.8}
\end{equation} 
from which it can be seen that the external source of the scalar field can effectively destroy the conservation of energy and momentum of the AdS system.

In this work, we shall focus on the case in which the system is inhomogeneous in a single spatial direction and is assumed to be periodic.
The metric ansatz is as follows
\begin{equation}
	ds^{2}=\Sigma^{2}\left(Gdx^{2}+G^{-1}dy^{2}\right)+2dt\left(dr-Adt-Fdx\right),\label{eq:2.9}
\end{equation}
where all metric fields are functions of $(t, x, r)$, and the translational invariance in the $y$-direction is preserved.
{In this case, the non-zero components of the energy-momentum tensor (\ref{eq:2.6a}) are $\left(\epsilon,\eta,p_{L},p_{T}\right) = \left(-T^{t}_{t},-T^{t}_{x},T^{x}_{x},T^{y}_{y}\right)$, which represent the energy density, momentum density, the longitudinal and transverse pressures of the system, respectively.
Since the system we consider is homogeneous in the $y$-direction and has no flow, there is no momentum in the $y$-direction and the shear force, so the corresponding energy-momentum tensor components are all zero $\left(T^{t}_{y},T^{x}_{y}\right)=\left(0,0\right)$.}
The non-trivial components of the Ward-Takahashi identity (\ref{eq:2.8}) reduce to
\begin{subequations}
	\begin{align}
		\partial_{t}\epsilon-\partial_{x}\eta&=-\left\langle O\right\rangle\partial_{t}\phi_{0},\label{eq:2.10a}\\
		\partial_{t}\eta-\partial_{x}p_{L}&=-\left\langle O\right\rangle\partial_{x}\phi_{0}.\label{eq:2.10b}
	\end{align}\label{eq:2.10}
\end{subequations}
Based on the second relationship above (\ref{eq:2.10b}), we expect that the local inhomogeneous scalar source can provide a damping mechanism for the moving system to evolve from a steady flow to a static state.

\subsection{Phase diagram}

\begin{figure}
	\begin{center}
		\subfigure[]{\includegraphics[width=.49\linewidth]{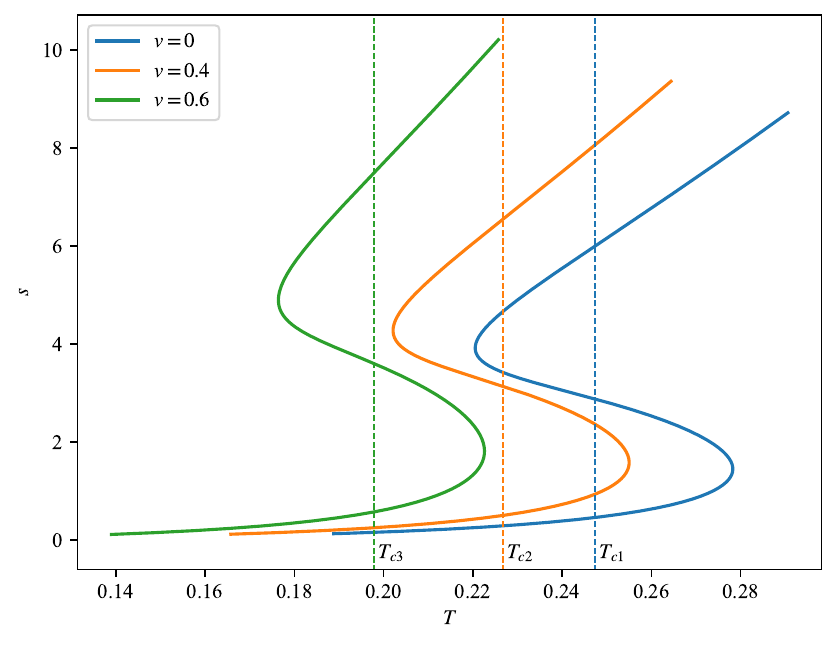}\label{fig:1}}
		\subfigure[]{\includegraphics[width=.49\linewidth]{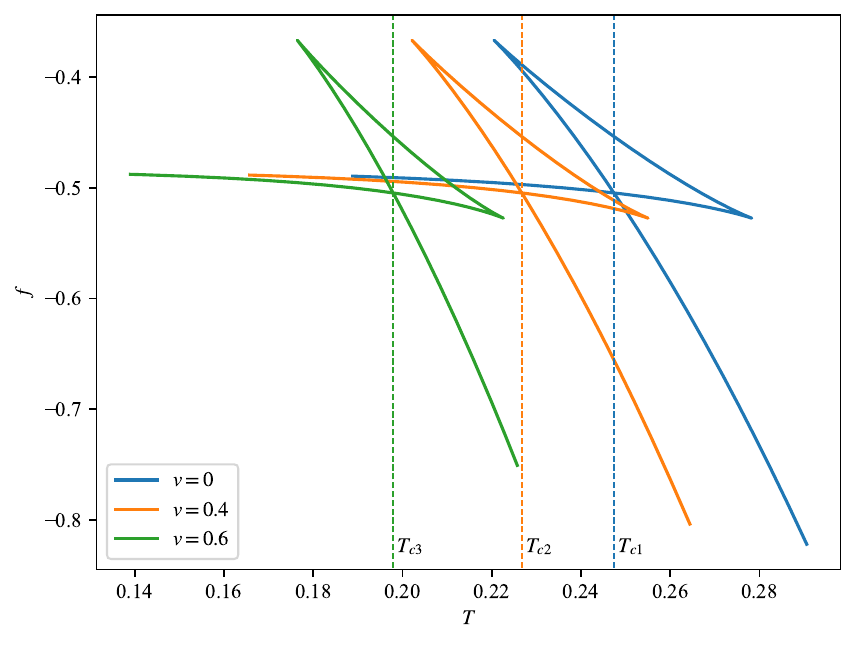}\label{fig:2}}
		\subfigure[]{\includegraphics[width=.49\linewidth]{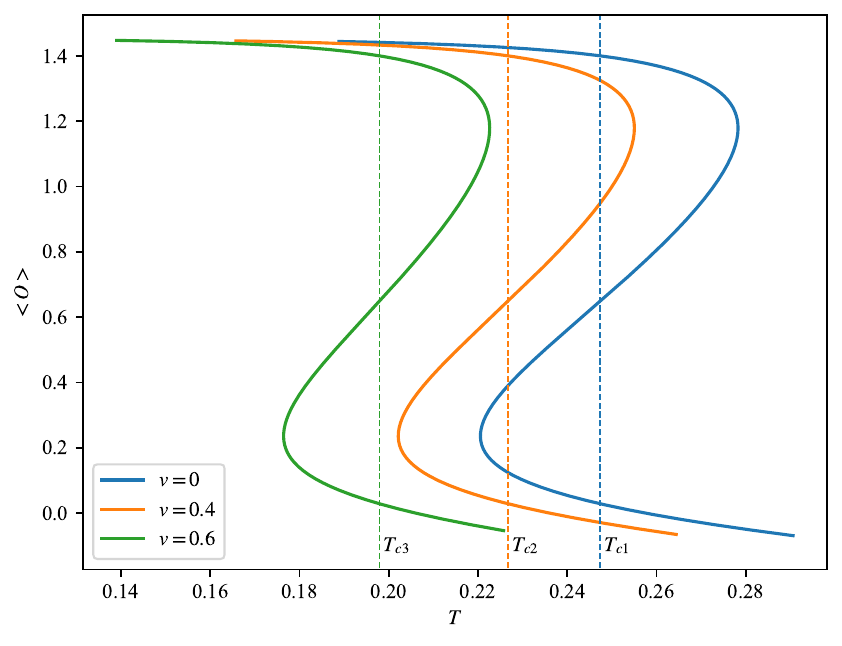}\label{fig:3}}
		\subfigure[]{\includegraphics[width=.49\linewidth]{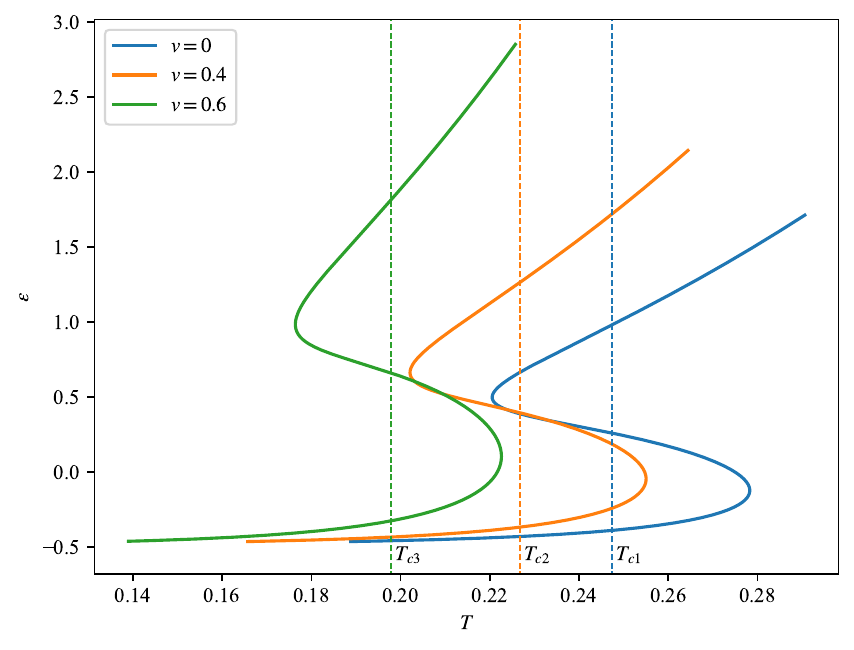}\label{fig:4}}
		\subfigure[]{\includegraphics[width=.49\linewidth]{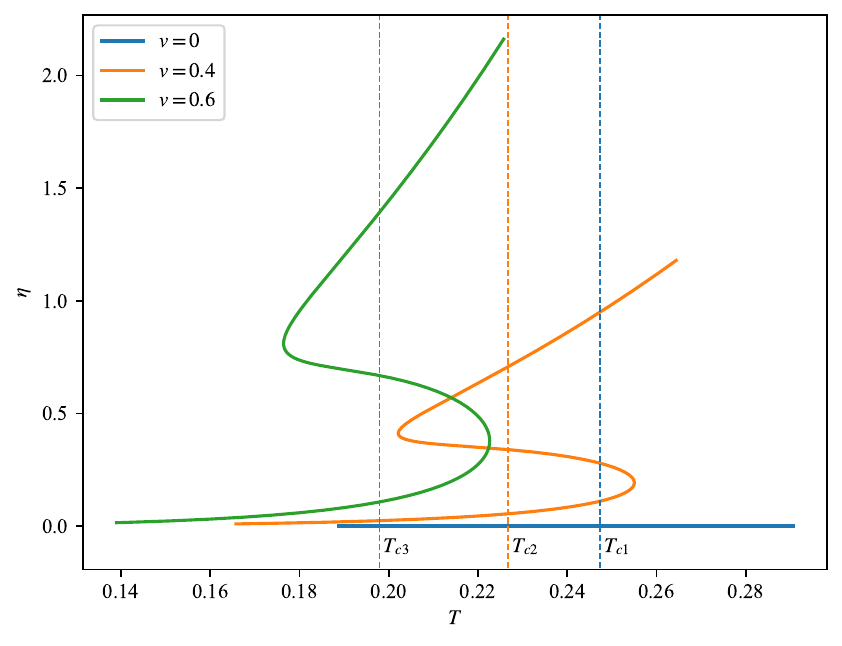}\label{fig:5}}
		\subfigure[]{\includegraphics[width=.49\linewidth]{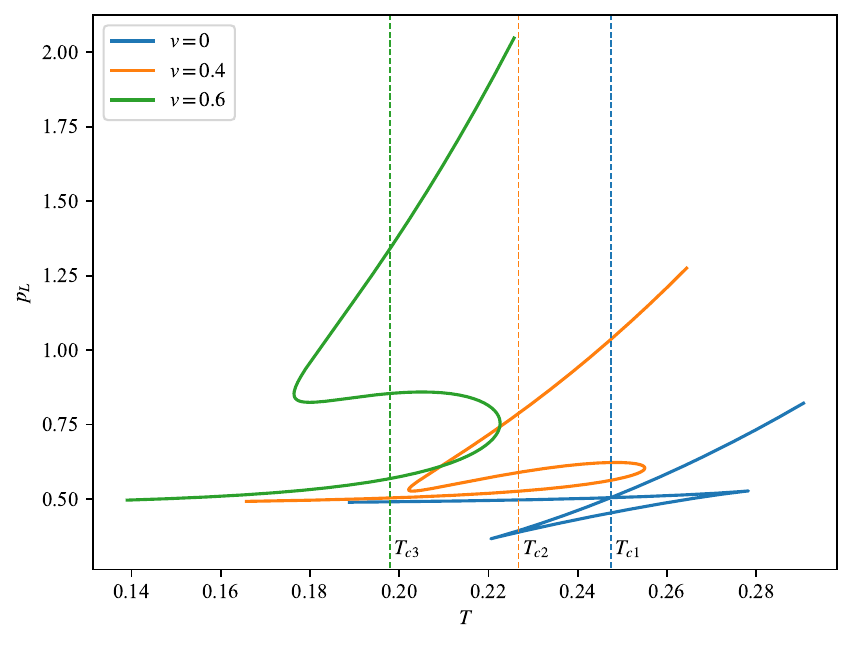}\label{fig:6}}
		\caption{Entropy density $s$, free energy density $f$, the expectation value of the scalar operator $\left\langle O\right\rangle$, energy density $\epsilon$, momentum density $\eta$ and longitudinal pressure $p_{L}$ versus temperature $T$. The vertical dotted line represents the phase transition temperature $T_{c}$. Curves of different colors represent different flow velocities $v$. Due to the thermodynamic relationship $p_{T}=-f$, we do not show the transverse pressure of the system.}\label{fig:1-6}
	\end{center}
\end{figure}

Previous works on holographic first-order phase transitions have only considered the phase diagram of a static system.
In that case, one sets $F=0$ and $G=1$, resulting in vanishing momentum density $\eta=0$ and equal longitudinal and transverse pressures $p_{L}=p_{T}$.
The system can be parameterized by a single physical quantity, such as energy density.

In order to obtain the steady flow with non-zero momentum density, we need to abandon such an assumption and retain all components of the metric ansatz (\ref{eq:2.9}).
The system at this time has two independent physical degrees of freedom.
Note that in this section we are concerned with the time-independent homogeneous solution, where the field functions depend only on the coordinate $r$, and set the scalar source to $\phi_{0}=1$.
{After obtaining the geometric configuration and setting the effective Newton constant $\kappa^{2}_{4}=8\pi G_{4}=1$ with the four-dimensional Newton constant $G_{4}$, following black hole thermodynamics, one can extract the temperature and entropy density of the system from the surface gravity and area element of the black hole horizon respectively, as follows}
\begin{equation}
	T=\frac{1}{2\pi}\frac{d}{dr}\left[A+\frac{F^{2}}{2G\Sigma^{2}}\right]_{r=r_{h}},\quad s=2\pi\Sigma^{2}(r_{h}),
\end{equation}
where $r_{h}$ represents the location of the event horizon.
{For our case without the additional Noether flow, the free energy density satisfies the following thermodynamic relation \cite{Gursoy:2008za}}
\begin{equation}
	f=\epsilon-sT-v\eta,
\end{equation}
where the flow velocity $v$ can be extracted from the unit eigenvector of the energy-momentum tensor $\left\langle T^{i}_{j}\right\rangle$.

Figure \ref{fig:1-6} shows the phase diagrams under different flow velocities, which are related through the usual boost transformation.
For a steady flow with flow velocity $v$, after applying the Lorentz transformation
\begin{equation}
	t=\gamma\left(t'-vx'\right),\quad x=\gamma\left(x'-vt'\right),\quad \gamma=\frac{1}{\sqrt{1-v^{2}}},\label{eq:2.13}
\end{equation}
the geometric configuration (\ref{eq:2.9}) degenerates into the static situation
\begin{equation}
	ds^{2}=\Sigma'^{2}\left(dx'^{2}+dy^{2}\right)-2A'dt'^{2}-2\omega_{i}dx^{i}dr, \label{eq:2.14}
\end{equation}
with a boundary one-form $\omega=\gamma\left(-dt'+vdx'\right)$.
Such a metric form describes a comoving fluid system relative to an observer whose $3$-velocity components equal $\omega^{i}$ \cite{Chesler:2013lia}.
To this moving observer, the fluid is at rest.
Note that in the frame (\ref{eq:2.9}), the boundary one-form is $-dt$, indicating a stationary observer.
Our numerical results show that for any observer, the relatively stationary system has the same phase structure, that is, the case of $v=0$ in figure \ref{fig:1-6}.
In other words, in the frame (\ref{eq:2.14}), the phase structure of the system does not depend on the parameter $v$.

For a stationary observer, there are significant thermodynamic differences between a static fliud and a steady flow.  
Qualitatively, this difference is reflected in the momentum density and longitudinal pressure. 
As can be seen in figure \ref{fig:5}, in sharp contrast to the vanishing momentum density of a static fluid, the phase structure of the momentum density of a steady flow is similar to that of the energy density, showing an S-shape.
In this case, the momentum densities of the high-energy phase and the low-energy phase at the phase transition temperature are not equal, resulting in the inability of these two thermal phases to coexist statically.
Such a momentum density gradient will induce an energy flow, causing the high-energy phase with greater momentum density to move in the low-energy phase.
In particular, we find the following relationship
\begin{equation}
	\eta_{\text{high}}(T_{c})-\eta_{\text{low}}(T_{c})=v\left(\epsilon_{\text{high}}(T_{c})-\epsilon_{\text{low}}(T_{c})\right),\label{eq:2.15}
\end{equation}
which, combined with equation (\ref{eq:2.10a}), states that the domain wall connecting the coexisting phases will move with velocity $v$.
Note that the quantity in parentheses on the right side of the equation (\ref{eq:2.15}) is generally defined as the latent heat of a first-order phase transition $H=\Delta\epsilon(T_{c})$.
On the other hand, for a static fluid, due to the equivalence of the spatial $x$-direction and the spatial $y$-direction, the longitudinal pressure is equal to the transverse pressure, satisfying $p_{L}=p_{T}=-f$.
Like the free energy density, the phase structure of the longitudinal pressure of the static fluid exhibits a swallowtail shape.
Such a structure indicates that the longitudinal pressures of the coexisting phases are equal, ensuring the dynamic equilibrium of the domain wall.
Differently, for the steady flow, the non-zero momentum density breaks the inversion symmetry in the $x$-direction of space, resulting in inconsistency between the longitudinal pressure and the transverse pressure.
As shown in figure \ref{fig:6}, the phase structure of the longitudinal pressure of the system changes from the swallowtail shape to the S-shape with the increase of the flow velocity, indicating that a pressure difference gradually forms between the coexisting phases inside and outside the domain wall.
Furthermore, we find that the pressure difference satisfies
\begin{equation}
	\left(p_{L}\right)_{\text{high}}(T_{c})-\left(p_{L}\right)_{\text{low}}(T_{c})=v\left(\eta_{\text{high}}(T_{c})-\eta_{\text{low}}(T_{c})\right),\label{eq:2.16}
\end{equation}
which, combined with equation (\ref{eq:2.10b}), shows that the domain wall can stably move at velocity $v$.
Combined with relationship (\ref{eq:2.15}), on the other hand, one can also express such a  pressure difference in terms of latent heat
\begin{equation}
	\Delta p_{L}(T_{c})=v^{2}H.\label{eq:2.17}
\end{equation}
{Recently, simulating cosmological phase transitions via nucleation of bubbles in a holographic first-order phase transition model, the velocity of planar domain walls during nucleation induced by a supercooled state is revealed \cite{Bea:2021zsu}. 
The numerical simulation results show that the velocity of the domain wall depends linearly on the pressure difference between the inside and outside of the bubble, namely $\Delta p\sim v$. This is a non-equilibrium situation, in which the bubble is growing. As a complement, for a steady-flowing bubble, the gap (\ref{eq:2.17}) indicates that the velocity of the domain wall has a different dependence on the pressure difference, expressed as $\Delta p \sim v^{2}$.
Such results indicate that the dependence of domain wall velocity on the pressure difference between the inside and outside of the bubble depends on the state of its motion.
On the other hand, the topological structure of the domain wall is also an important factor affecting such dependence \cite{Li:2023xto}.}

\begin{figure}
	\begin{center}
		\subfigure[]{\includegraphics[width=.49\linewidth]{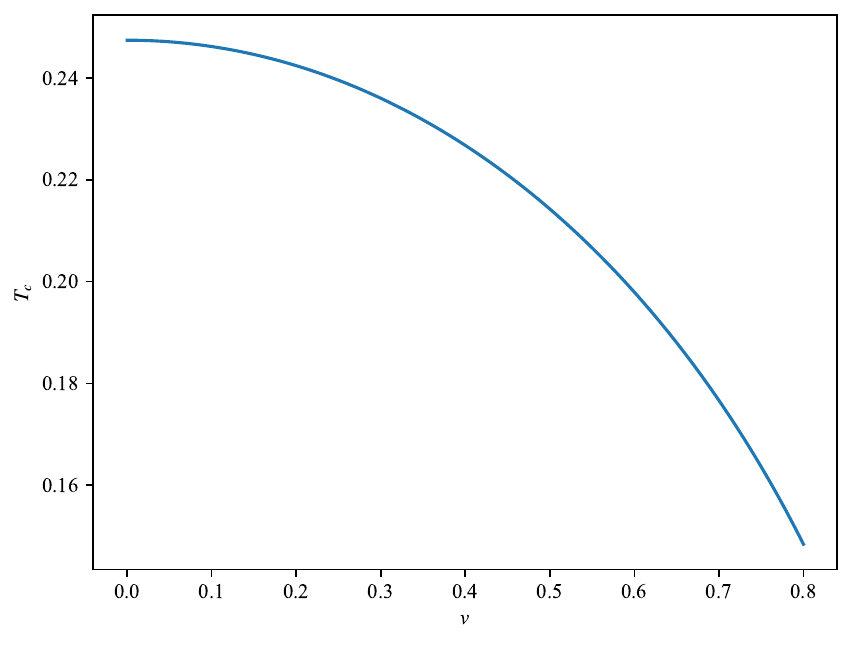}\label{fig:8}}
		\subfigure[]{\includegraphics[width=.49\linewidth]{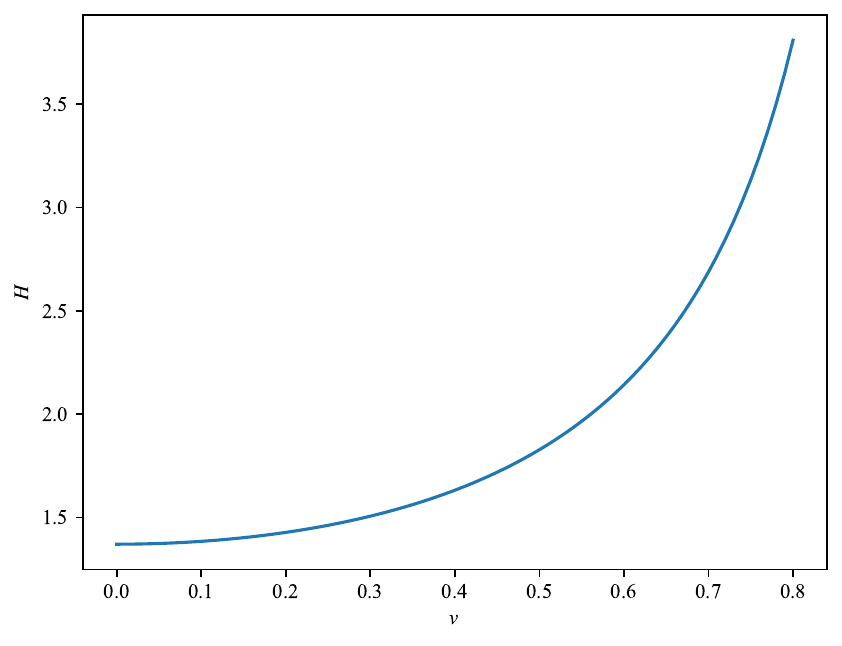}\label{fig:9}}
		\caption{Phase transition temperature $T_{c}$ and latent heat $H$ versus flow velocity $v$.}\label{fig:8-9}
	\end{center}
\end{figure}

Quantitatively, figure \ref{fig:8-9} shows the phase transition temperature and latent heat of the first-order phase transition as a function of the flow velocity.
The monotonically decreasing curve in figure \ref{fig:8} indicates that a steady flow with a faster flow velocity has a lower phase transition temperature.
Such a result shows that for a fast-flowing fluid, the first-order phase transition is difficult to occur.
On the other hand, as shown in figure \ref{fig:9}, the latent heat of phase transition increases monotonically with the increase of flow velocity, indicating that the faster the fluid flows, the more energy needs to be absorbed or released to transform between different thermal phases.
From relationship (\ref{eq:2.17}), the increasing latent heat ensures that the pressure difference between the coexisting phases increases with the increase of flow velocity, providing sufficient driving force for the movement of the domain wall.
{An interesting question is what happens when the flow velocity approaches the speed of light.
Unfortunately, due to numerical problems, we cannot obtain the geometric configuration for the case $v>0.8$, leaving this question unresolved and awaiting further exploration in the further. However, from the trend of the curves in Figure \ref{fig:8-9}, we can guess that a possible result is that when the flow velocity approaches the speed of light, the phase transition temperature approaches zero and the latent heat tends to infinity, leading to the disappearance of the first-order phase transition. In order to verify this speculation, a possible strategy is to use an analytical approximation to solve the static field equations to extract the physical quantities in the extreme case.}

\section{Phase separation}\label{sec:Ps}
For the static system, previous works have verified that the thermal phases with equal free energy density at the phase transition temperature can coexist stably through a domain wall connection, forming a spatially inhomogeneous phase-separated state.
Similarly, for the steady flow, picture \ref{fig:2} shows that the two thermodynamically stable phases at the phase transition temperature have equal free energy density and are therefore expected to coexist.
In this section, through fully nonlinear numerical simulations of the dual gravitational dynamics, we investigate the real-time dynamics of the steady flow near a first-order phase transition and reveal the phase-separated state with non-zero momentum.
Depending on the dynamical instability of the steady flow, there are two types of dynamical transitions between the homogeneous and inhomogeneous states, which will be introduced in the following two subsections respectively.
Without loss of generality, we choose the steady flow system with flow velocity $v=0.4$.

\subsection{Spinodal decomposition}\label{sec:up}
For the spinodal region where the specific heat is negative, the thermodynamic instability of the system induces a dynamical instability \cite{Janik:2015iry,Janik:2016btb}.
At the linear level, there is a branch of hydrodynamic modes where the imaginary part of the frequency is positive in the long wavelength region.
Under arbitrarily small perturbations, such unstable modes will grow exponentially and push the system away from the initial equilibrium state, inducing a dynamical transition. 

\begin{figure}
	\begin{center}
		\subfigure[]{\includegraphics[width=.49\linewidth]{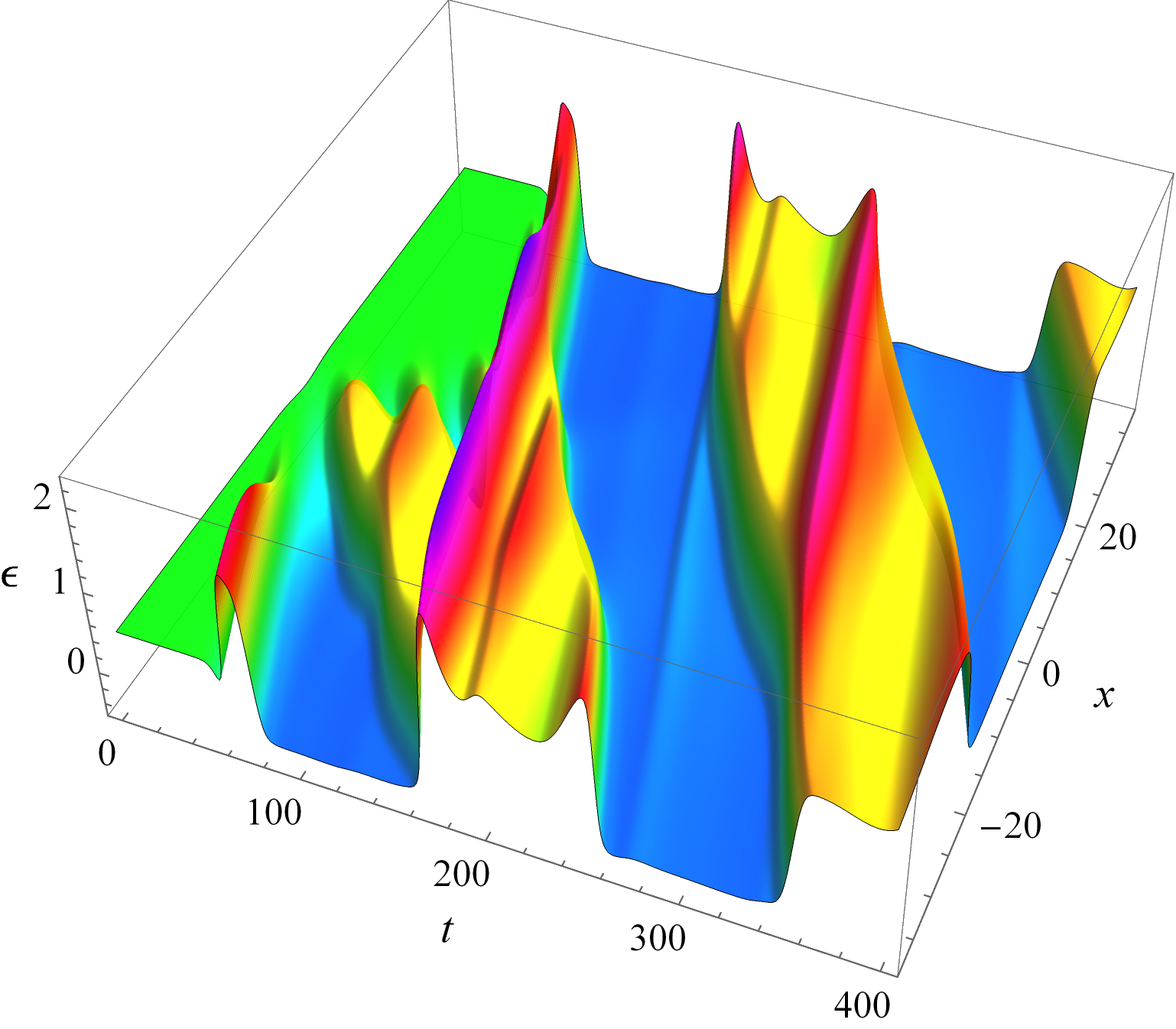}\label{fig:10}}
		\subfigure[]{\includegraphics[width=.49\linewidth]{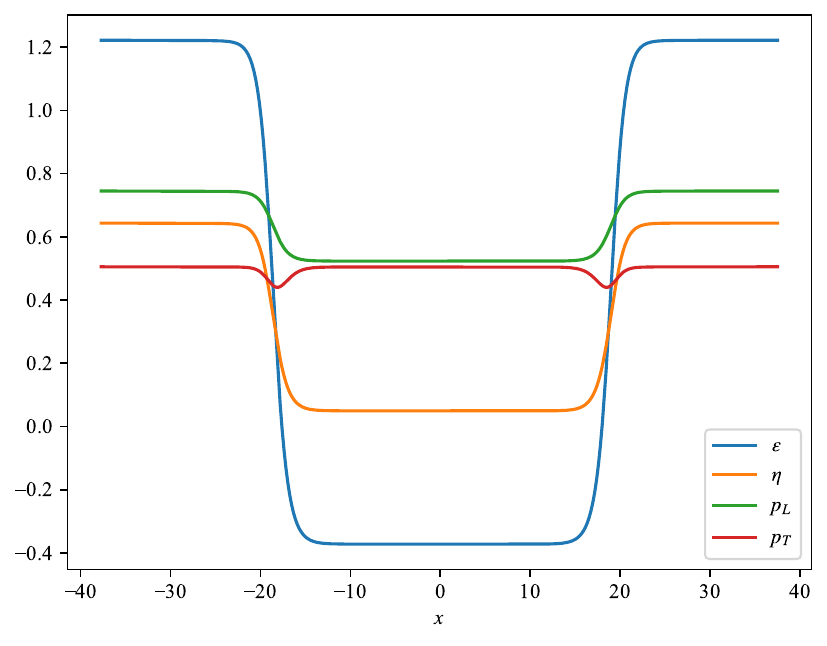}\label{fig:11}}
		\caption{(a) Real-time dynamics of the energy density of the system. The initial data is a steady flow with velocity $v=0.4$ and energy density $\epsilon=0.43$, located in the spinodal region, plus a small perturbation. (b) The spatial distribution of the physical quantities in the final state.}
		\label{fig:10-11}
	\end{center}
\end{figure}

In order to reveal such a dynamical process in the case of steady flow, we select a steady flow in the spinodal region as the initial data and then apply a small perturbation.
Figure \ref{fig:10} shows the real-time dynamics of the energy density of the system, from which it can be seen that a dynamical transition occurs that destroys the spatial translational symmetry.
The entire dynamical process can be roughly divided into three stages.
Initially, several energy peaks are spontaneously excited from the homogeneous geometric configuration and move along the flow direction of the background.
Subsequently, these energy peaks merge together to form a moving domain, accompanied by fluctuations in energy density.
Finally, as the energy fluctuations decay, the system gradually settles down to a phase-separated state with non-zero momentum.
During this evolution, from the Ward-Takahashi identity (\ref{eq:2.10}), the total energy and total momentum of the system are conserved.

The final state of the evolution consists of coexisting phases at the phase transition temperature, which occupy different spatial regions and are connected by domain walls.
The conservation of total momentum results in the system not being at rest, but flowing steadily at a uniform velocity as in the initial state.
Since the momentum density of the low-energy phase is close to zero, as shown in figure \ref{fig:5}, such a geometric configuration can be approximated as a dynamic structure of a moving domain of the high-energy phase in the background of a static low-energy bath.
Note that the velocity of the moving domain in the final state is $v=0.37$, compared with the velocity $v=0.4$ of the initial steady flow, indicating that the phase transition reduces the flow velocity of the system.
For such a time-dependent inhomogeneous {stationary state} with uniform flow velocity $v$, there exists a Killing vector
\begin{equation}
	\vartheta=\partial_{t}-v\partial_{x}.\label{eq:3.1}
\end{equation}
Combined with the Ward-Takahashi identity (\ref{eq:2.10}), one can conclude that the spatial distribution of the momentum density and longitudinal pressure in the final state satisfies the following relationship
\begin{subequations}
	\begin{align}
		\eta(t,x)&=v\epsilon(t,x)+C_{1},\\
		p_{L}(t,x)&=v\eta(t,x)+C_{2},
	\end{align}
\end{subequations}
where $C_{1}$ and $C_{2}$ are both constants, as shown in figure \ref{fig:11}.
Such results once again verify the relationship between the momentum density gradient between the coexisting phases, longitudinal pressure difference and the latent heat, (\ref{eq:2.15}) and (\ref{eq:2.16}).
As for the transverse pressure, as mentioned before, it is equal to the free energy density for a thermal phase and is therefore a constant in the phase domain except for the domain walls.

\subsection{Condensation nucleus}\label{sec:mp}
For the thermodynamically metastable region adjacent to the spinodal region, the system is dynamically stable at the linear level due to the absence of unstable modes, indicating that small perturbations will be dissipated and are not sufficient to change the state of the system.
However, inspired by the fact that condensation nuclei induce phase transitions of supercooled water in reality, the thermodynamically metastable state is verified to suffer from a type of nonlinear instability \cite{Bea:2020ees,Bea:2022mfb,Chen:2022cwi}.
In this case, the system is in a local ground state.
The occurrence of a phase transition requires a disturbance of sufficient intensity to push the system across a dynamical barrier and thereby evolve to another local ground state.
Near the threshold of the disturbance intensity, a critical state will appear.

\begin{figure}
	\begin{center}
		\subfigure[]{\includegraphics[width=.49\linewidth]{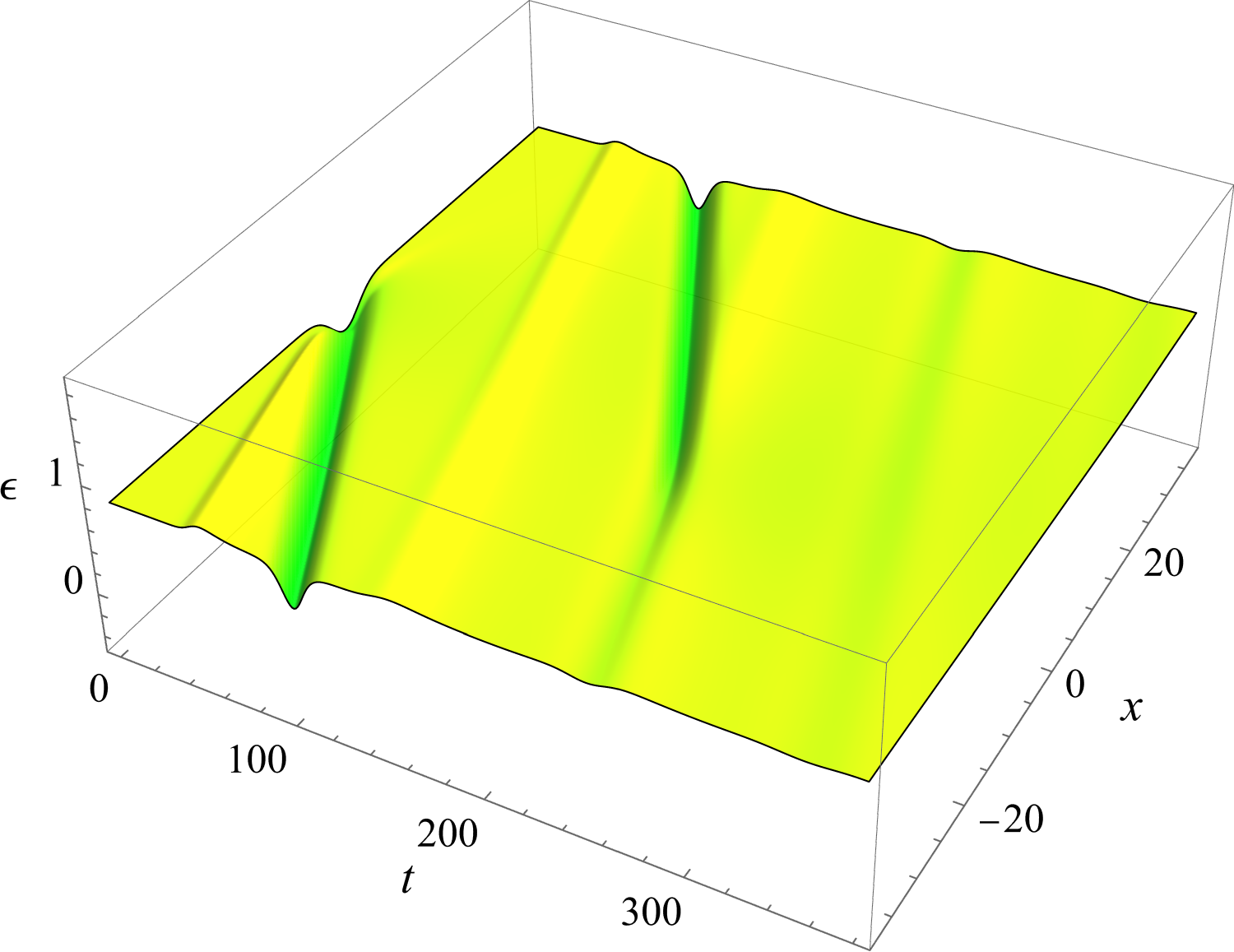}}
		\subfigure[]{\includegraphics[width=.49\linewidth]{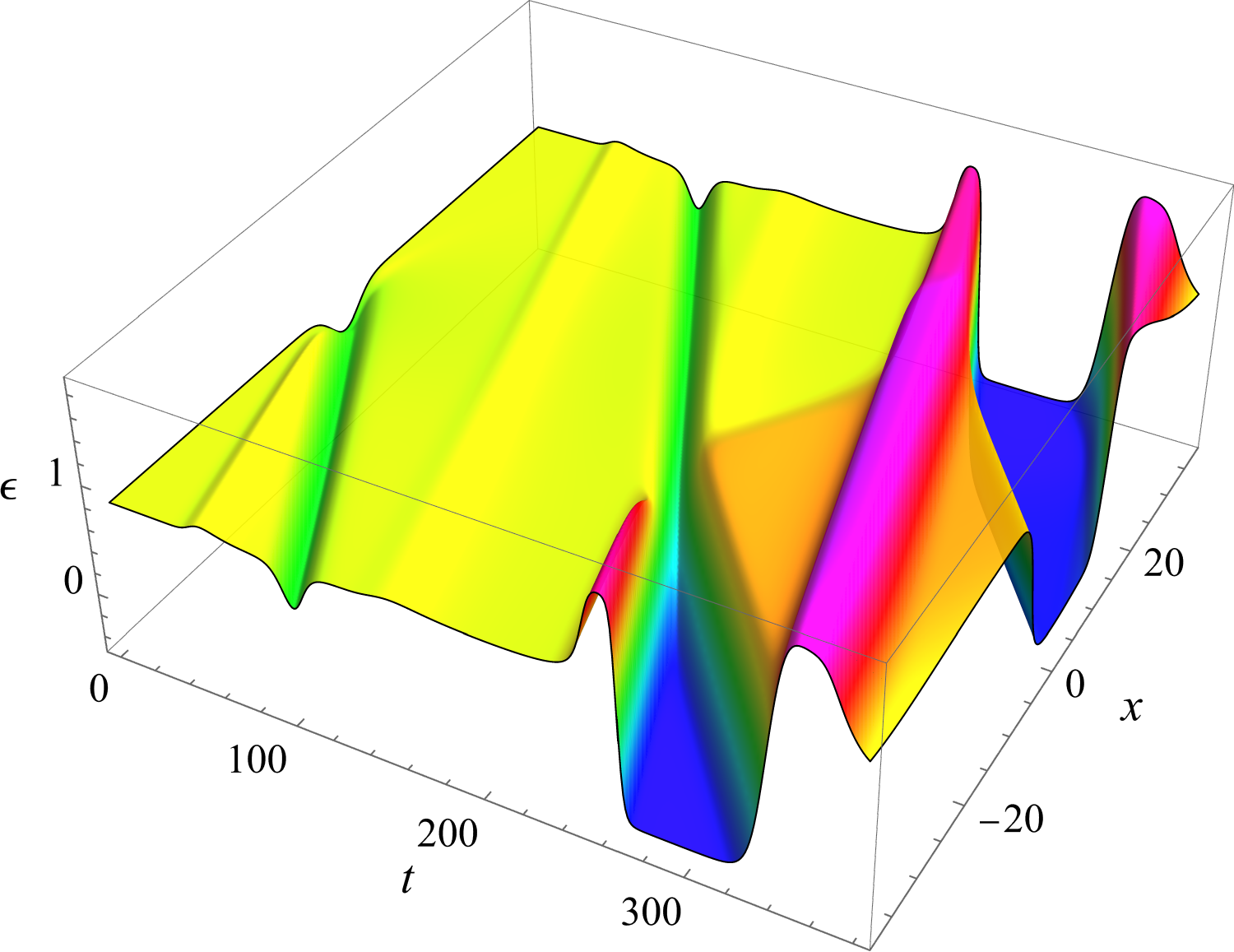}}
		\caption{Real-time dynamics of the energy density of the system in the case of subcritical disturbance (a) and supercritical disturbance (b). The initial data is a steady flow with velocity $v=0.4$ and energy density $\epsilon=0.81$, located in the supercooled region, plus a sufficiently large disturbance.}\label{fig:12-13}
	\end{center}
\end{figure}

In order to reveal the nonlinear instability of metastable steady flow, we choose a steady flow in the supercooled region as the initial data and then apply a disturbance parameterized by intensity.
Similarly, there exists a threshold that triggers the phase transition.
Figure \ref{fig:12-13} shows the real-time dynamics of the energy density of the system near the threshold of the disturbance intensity, from which it can be observed that, unlike the case of spinodal decomposition, a critical state appears in the dynamical intermediate process, and the final state of the evolution depends on the relative size of the actual disturbance intensity and the threshold.
The entire dynamical process is similar to the phase transition of supercooled water induced by condensation nuclei.
The initial disturbance needs to generate a local low-energy region that acts as a seed nucleus.
When the disturbance intensity approaches the threshold, the seed nucleus will reshape and quickly converge to a critical nucleus configuration, which moves along the flow direction of the background due to non-zero momentum.
The closer the disturbance intensity is to the threshold, the longer the critical nucleus exists in the dynamical process, exhibiting a critical behavior \cite{Chen:2022cwi}.
Finally, for the case where the disturbance intensity is slightly less than the threshold, the critical nucleus melts into the background and the system evolves back to a homogeneous steady flow.
On the contrary, for the case where the disturbance intensity is slightly greater than the threshold, the critical nucleus gradually grows and expands to form a low-energy phase, indicating that the system successfully crosses the dynamical barrier and transitions to another local ground state, that is, the phase-separated state with non-zero momentum.
Such results indicate that the metastable steady flow also suffers from the nonlinear instability, and there exists a moving critical nucleus that serves as a dynamical barrier to trigger the phase transition.

\section{Inhomogeneous source}\label{sec:Is}
In general, due to the confining boundary that prohibit matter from escaping, the {holographic dynamics essentially occurs in a closed system as its total energy is} conserved during the evolution.
In fact, such a conclusion depends on the boundary conditions of the field functions, that is, the external sources in (\ref{eq:2.4}), which are free parameters that cannot be determined by the field equations (\ref{eq:2.5}).
In our work, in order to ensure that the dual boundary system lives in a Minkowski spacetime, the external source of the metric fields is set to $\gamma_{0}=\text{diag}\{-1,1,1\}$.
On the other hand, the external source of the scalar field, which is an energy scale, is fixed to $\phi_{0}=1$.
In this case, it can be seen from the Ward-Takahashi identity (\ref{eq:2.10}) that the total energy and total momentum of the system are conserved in the dynamical process, providing a guarantee for obtaining a phase-separated state with non-zero momentum in a steady flow.

However, in reality, the system will inevitably interact with the external environment, resulting in the exchange of energy and momentum.
In this section, we simulate a stationary obstacle with a time-independent, locally inhomogeneous external source of the scalar field to reveal the real-time dynamics of steady flows and moving domains in the presence of interactions with the environment.
In particular, the scalar source is assumed to be a Gaussian distribution along the $x$-direction of space as follows
\begin{equation}
	\phi_{0}=1+pe^{-x^{2}/5},\label{eq:4.1}
\end{equation}
with amplitude $p$.
From the Ward-Takahashi identity (\ref{eq:2.10}), one can observe that such a scalar external source preserves the conservation of total energy, but induces the average momentum within a single period, defined as $J=\frac{1}{L_{x}}\int_{-L_{x}/2}^{L_{x}/2}dx\eta$ with period length $L_{x}$, to change as follows
\begin{equation}
	\frac{dJ}{dt}=-\frac{1}{L_{x}}\int_{-L_{x}/2}^{L_{x}/2}dx\left\langle O\right\rangle \frac{d\phi_{0}}{dx},
\end{equation}
from which{, together with the second law of thermodynamics,} we expect that the inhomogeneous scalar source can provide a damping mechanism for the moving system to decay from a {steady flow or a inhomogeneous stationary state} with $J\neq 0$ to a static state with $J=0$.

\subsection{For steady flow}
For the thermodynamically stable region with minimum free energy density, the system is dynamically stable at both the linear and nonlinear levels, indicating that it is in a global ground state and cannot undergo phase transitions.
In order to avoid phase transitions induced by the change of the external source, which can be regarded as a disturbance, we choose a steady flow in the thermodynamically stable high-energy region as the background and then impose the scalar source (\ref{eq:4.1}) to reveal the dynamical response of the system.

\begin{figure}
	\begin{center}
		\subfigure[]{\includegraphics[width=.49\linewidth]{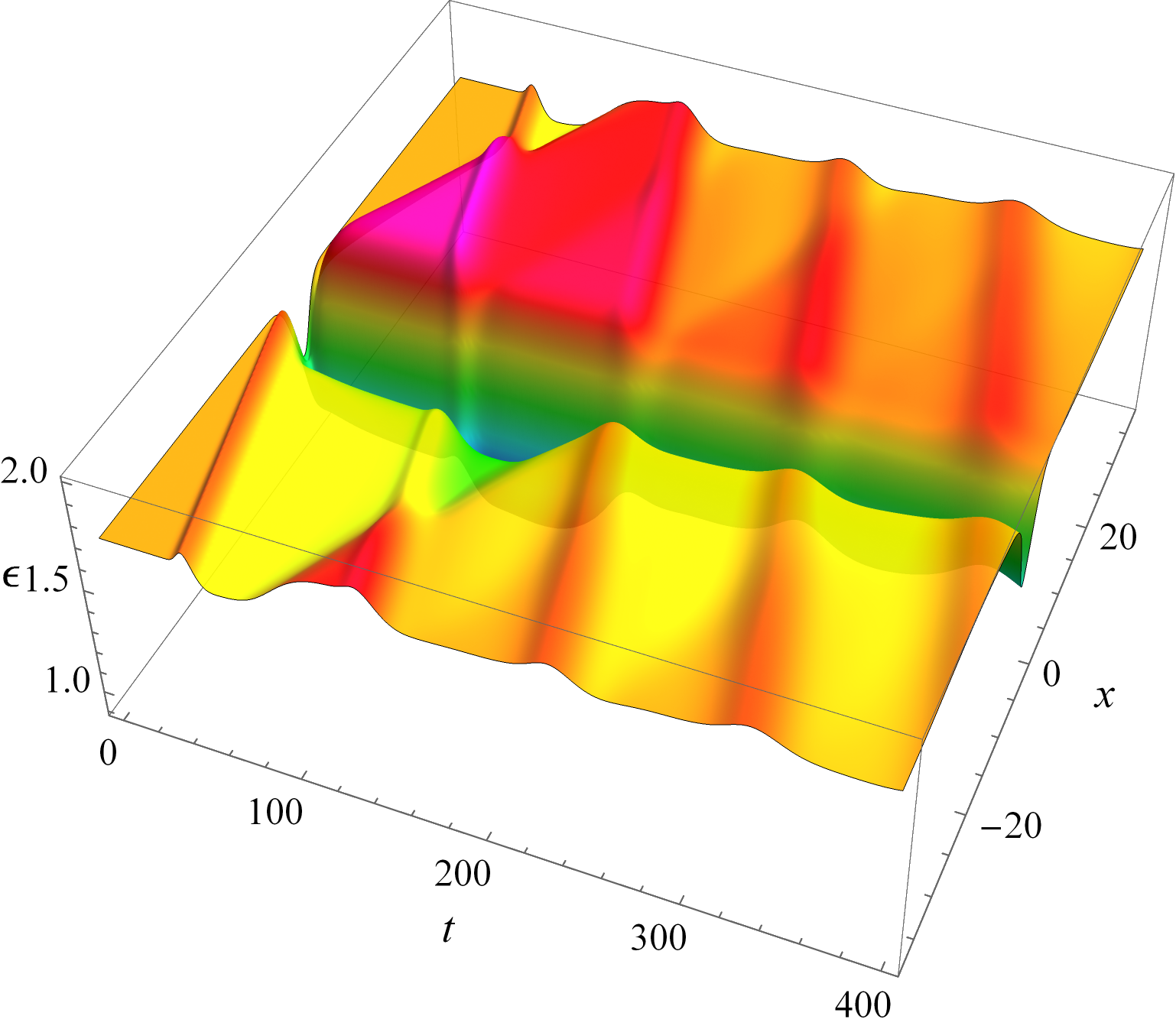}\label{fig:14}}
		\subfigure[]{\includegraphics[width=.49\linewidth]{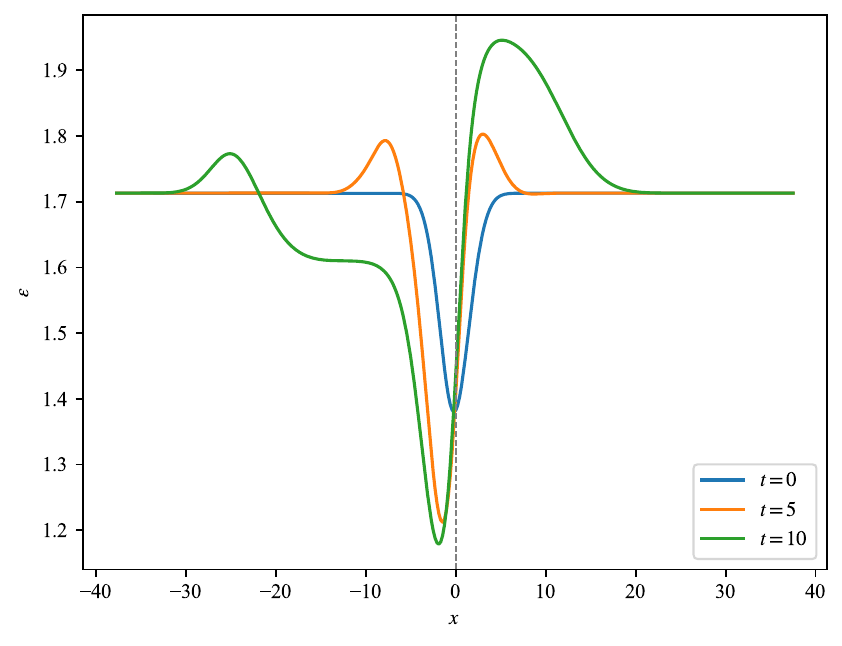}\label{fig:15}}
		\subfigure[]{\includegraphics[width=.49\linewidth]{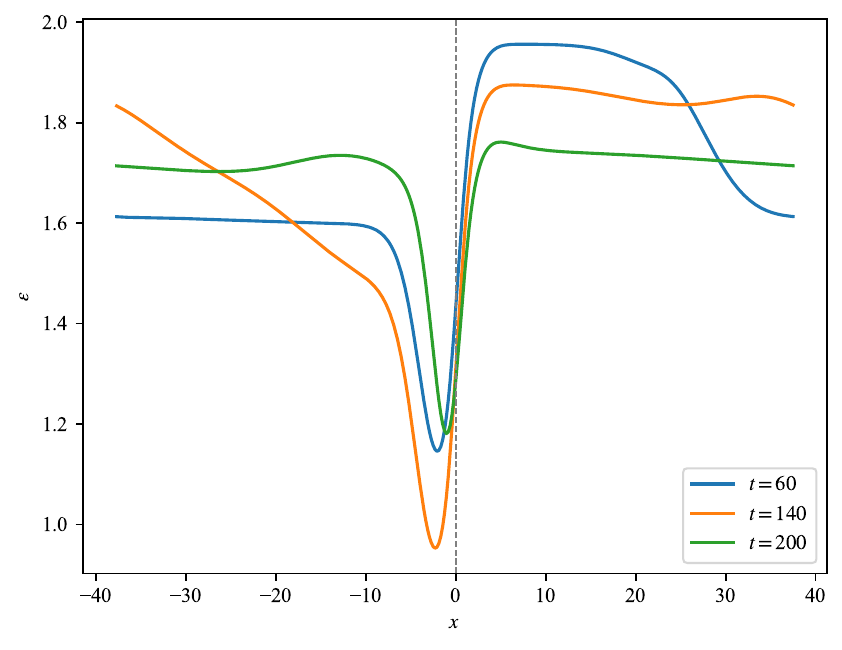}\label{fig:16}}
		\subfigure[]{\includegraphics[width=.49\linewidth]{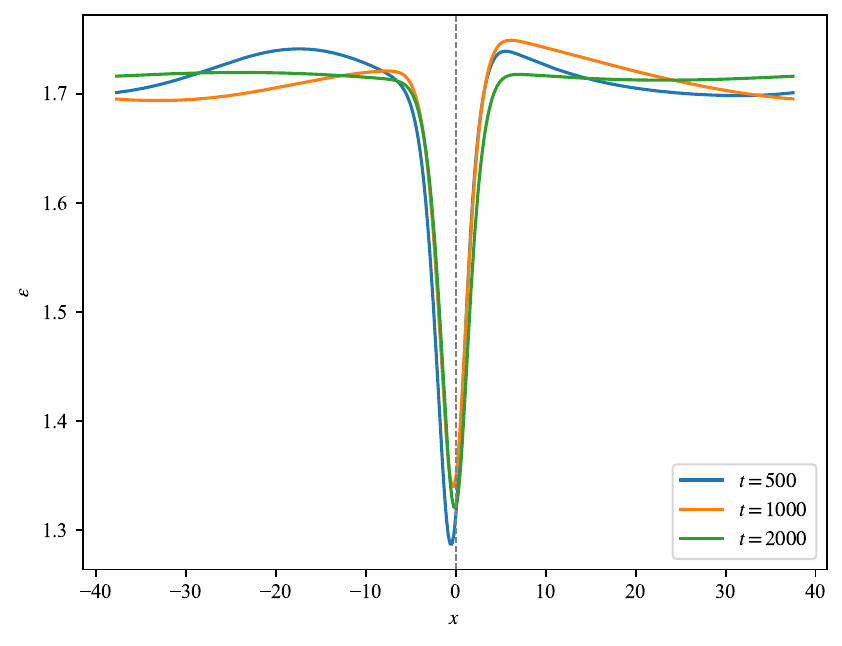}\label{fig:17}}
		\caption{Real-time dynamics of the energy density of a steady flow with velocity $v=0.4$ and energy density $\epsilon=1.71$, located in the thermodynamically stable high-energy region, under the action of an inhomogeneous scalar source (\ref{eq:4.1}) with amplitude $p=0.4$. Curves of different colors represent the spatial distribution of energy density at different times.}
		\label{fig:14-17}
	\end{center}
\end{figure}

Figure \ref{fig:14-17} shows the real-time dynamics of the energy density of the system, from which it can be seen that the inhomogeneous scalar external source breaks the dynamical equilibrium of the steady flow and induces an intrinsic transition.
The entire evolution process can be divided into three characteristic stages.
In the initial stage, as shown in figure \ref{fig:15}, this form of inhomogeneous scalar source (\ref{eq:4.1}) generates an energy well in the central region of the computational domain.
It is conceivable that such an energy well cannot maintain equilibrium in a steady flow.
The results of numerical simulation show that it migrates along the flow direction of the background and gradually deepens, accompanied by the excitation of energy peaks on both sides.
Since the flow velocity breaks the inversion symmetry of space, these two energy peaks exhibit completely distinct dynamical behaviors.
The energy peak on the left can propagate unimpeded along the flow direction, leaving a lower energy region along the way.
However, the movement of the energy peak on the right is hindered.
With the continuous injection of energy, it gradually grows and expands.
Such a dynamical process results in an energy gradient between the two sides of the obstacle.
At time $t=60$, the energy accumulated in the right area of the energy well reaches saturation, as shown in figure \ref{fig:16}.
During the second stage, the accumulated energy begins to spread outward.
Due to the periodic boundary condition, this energy will flow from the left boundary to the obstacle.
At time $t=140$, the energy well migrates to the greatest distance and reaches the maximum depth.
Subsequently, it gradually returns to the center with a decrease in depth and the disappearance of the energy gradient between the two sides.
In the final stage, as shown in figure \ref{fig:17}, as the energy fluctuations decay, the system stabilizes to a configuration of a homogeneous background plus a local energy well, just like the initial data.

\begin{figure}
	\begin{center}
		\subfigure[]{\includegraphics[width=.49\linewidth]{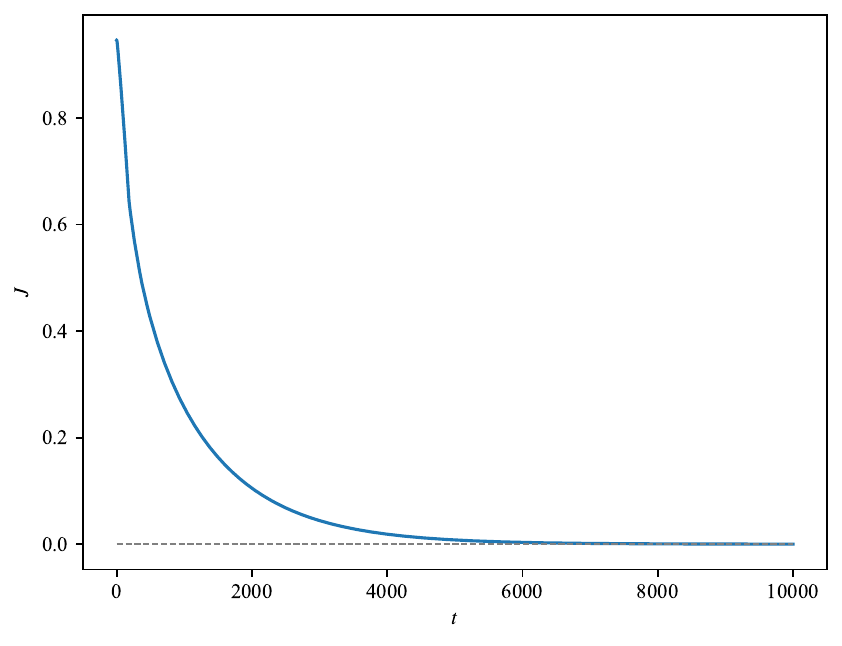}\label{fig:18}}
		\subfigure[]{\includegraphics[width=.49\linewidth]{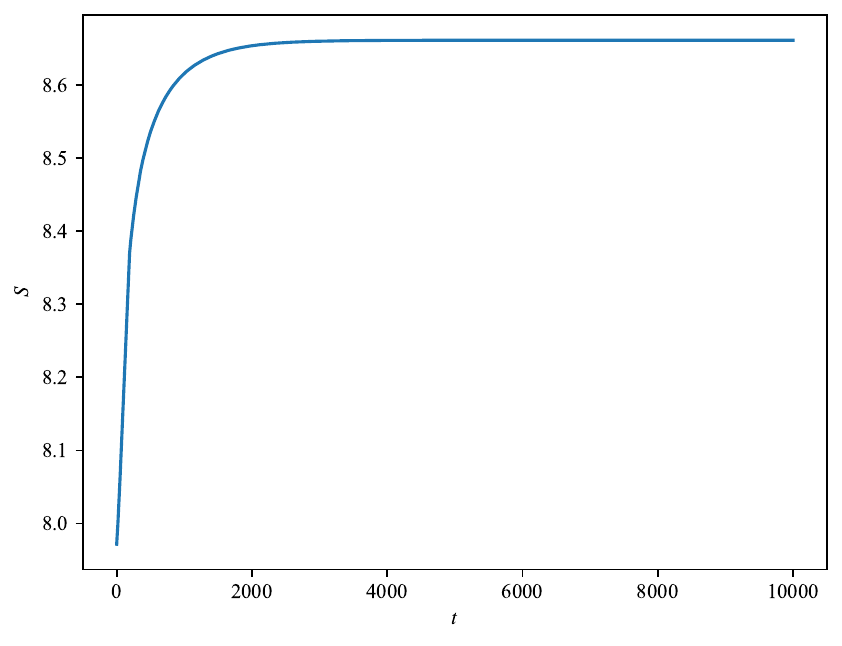}\label{fig:19}}
		\caption{The average momentum $J$ and average entropy, defined as  $S=\frac{1}{L_{x}}\int_{-L_{x}/2}^{L_{x}/2}dxs(r_{h})$ with apparent horizon radius $r_{h}$, of the system within a single period as a function of time during the dynamical process in figure \ref{fig:14-17}.}\label{fig:18-19}
	\end{center}
\end{figure}

Differently, the final state is a system at rest with zero momentum. 
As shown in figure \ref{fig:18}, the momentum of the system gradually decreases under the action of the inhomogeneous scalar source and eventually converges to zero, as expected.
During the dynamical process, the entropy increases monotonically, satisfying the second law of black hole mechanics, as shown in figure \ref{fig:19}.
Such results indicate that in a closed system with fixed energy, the system at rest has the largest entropy and is thus thermodynamically stabler relative to the moving system.

In fact, the final state must be a static configuration with zero flow velocity. 
This conclusion is universal and does not depend on the specific form of the scalar source.
{The reason is that the time-independent inhomogeneous source (\ref{eq:4.1}) makes the Killing vector (\ref{eq:3.1}) with non-zero velocity invalid, meaning $\vartheta[\phi_{0}]=-v\partial_{x}\phi_{0}\neq 0$.
For the final state of the evolution, the vector $\vartheta$ in the zero velovity case, namely $\partial_{t}$, is the only candidate for the Killing vector, indicating a static configuration.}
That is to say, under the action of a time-independent inhomogeneous obstacle, whose local structure destroys the symmetry of the solution space, the moving system is dynamically unstable and will inevitably lose momentum and degenerate into a static system.

\subsection{For moving domain}
In section \ref{sec:Ps}, we obtain the phase-separated configuration with non-zero momentum, which appears as a high-energy moving domain flowing steadily at a constant velocity in a low-energy bath.
If an inhomogeneous scalar source is imposed on such a steady-state geometry, there is no doubt that the moving domain will eventually cease and the system stabilizes to a static phase-separated configuration with a Killing vector $\partial_{t}$, as discussed previously.
Nonetheless, the dynamical intermediate processes still deserve further investigation, especially the dynamical behavior of the moving domain under the action of an inhomogeneous scalar source.
We find two characteristic dynamical patterns that the moving domains with different flow velocities exhibit when interacting with the obstacle.

For the initial data, we use the phase-separated configuration with non-zero momentum obtained through spinodal decomposition as the background and then impose the inhomogeneous scalar source (\ref{eq:4.1}) at the center of the low-energy phase. 
Similar to the previous subsection, such an operation generates an energy well in the central region of the low-energy bath as a potential barrier.
After a period of evolution, the moving high-energy domain will collide with it, allowing us to reveal the dynamical behavior of the moving domain when interacting with a fixed obstacle.
As expected, numerical results show that such a dynamical process depends on the flow velocity of the moving domain and the amplitude $p$ of the external source, which characterizes the strength of the potential barrier.
There are two dynamical patterns exhibited by the moving domain.
For the case of low velocity, as the amplitude $p$ increases, the moving domain exhibits passing, pinning and rebounding behaviors in sequence. When flowing at a high velocity, there is an intermediate range of the amplitude $p$ that causes the moving domain to split.
To this end, without loss of generality, we select a steady flow each in the spinodal region on the branches with flow velocities $v=0.2$ and $0.4$.
The selection rule for energy density requires that these steady flows with different velocities degenerate into the same static fluid with energy density $\epsilon=0.295$, located in the spinodal region of the static branch with $v=0$, after the corresponding Lorentz transformation.
After perturbations, the velocities of the resulting moving domains are $v=0.18$ and $0.37$ respectively.
For each case, we continuously adjust the amplitude $p$ to reveal the dynamical properties of the moving domain under interaction with obstacles of different strengths.

\begin{figure}
	\begin{center}
		\subfigure[]{\includegraphics[width=.49\linewidth]{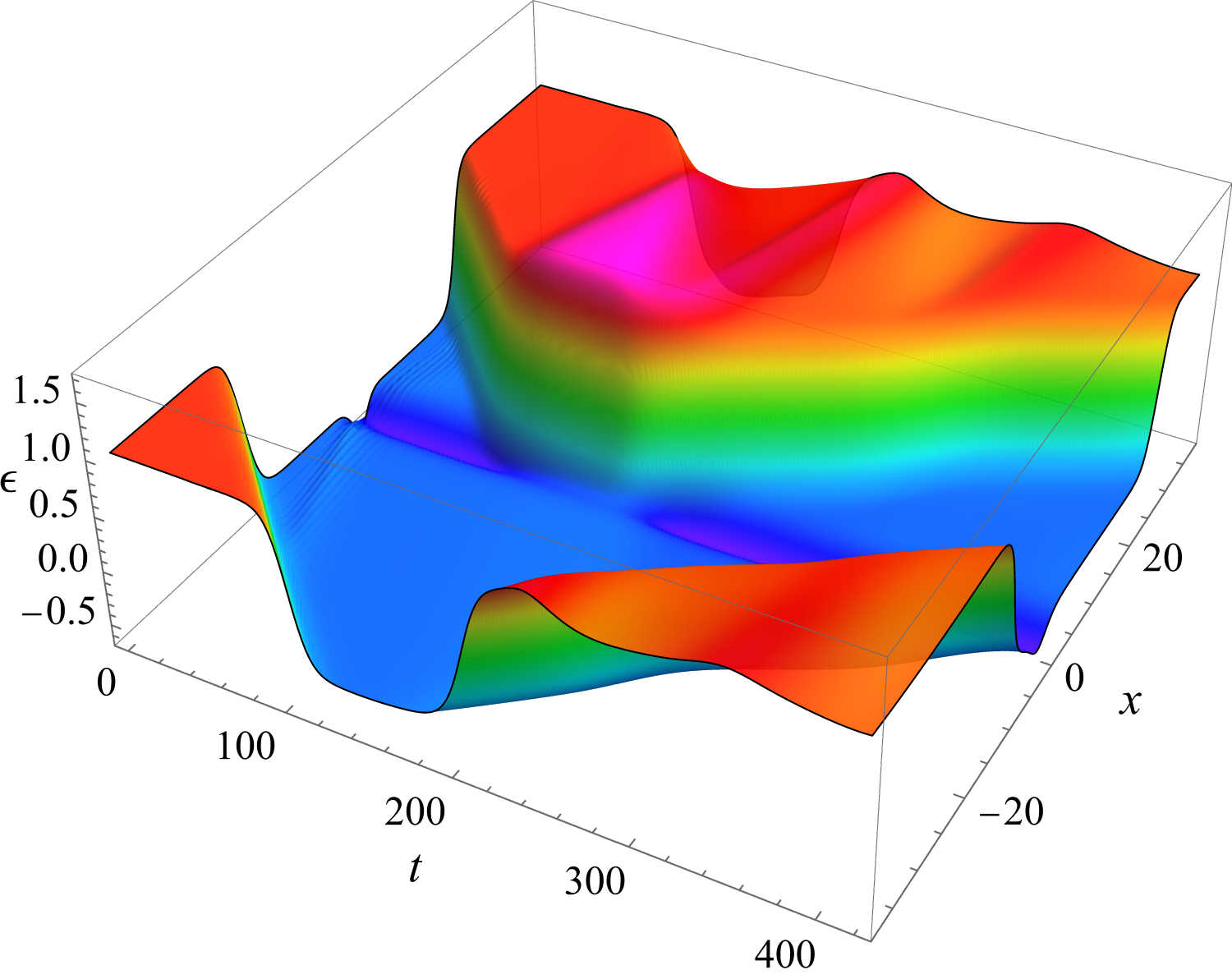}\label{fig:20}}
		\subfigure[]{\includegraphics[width=.49\linewidth]{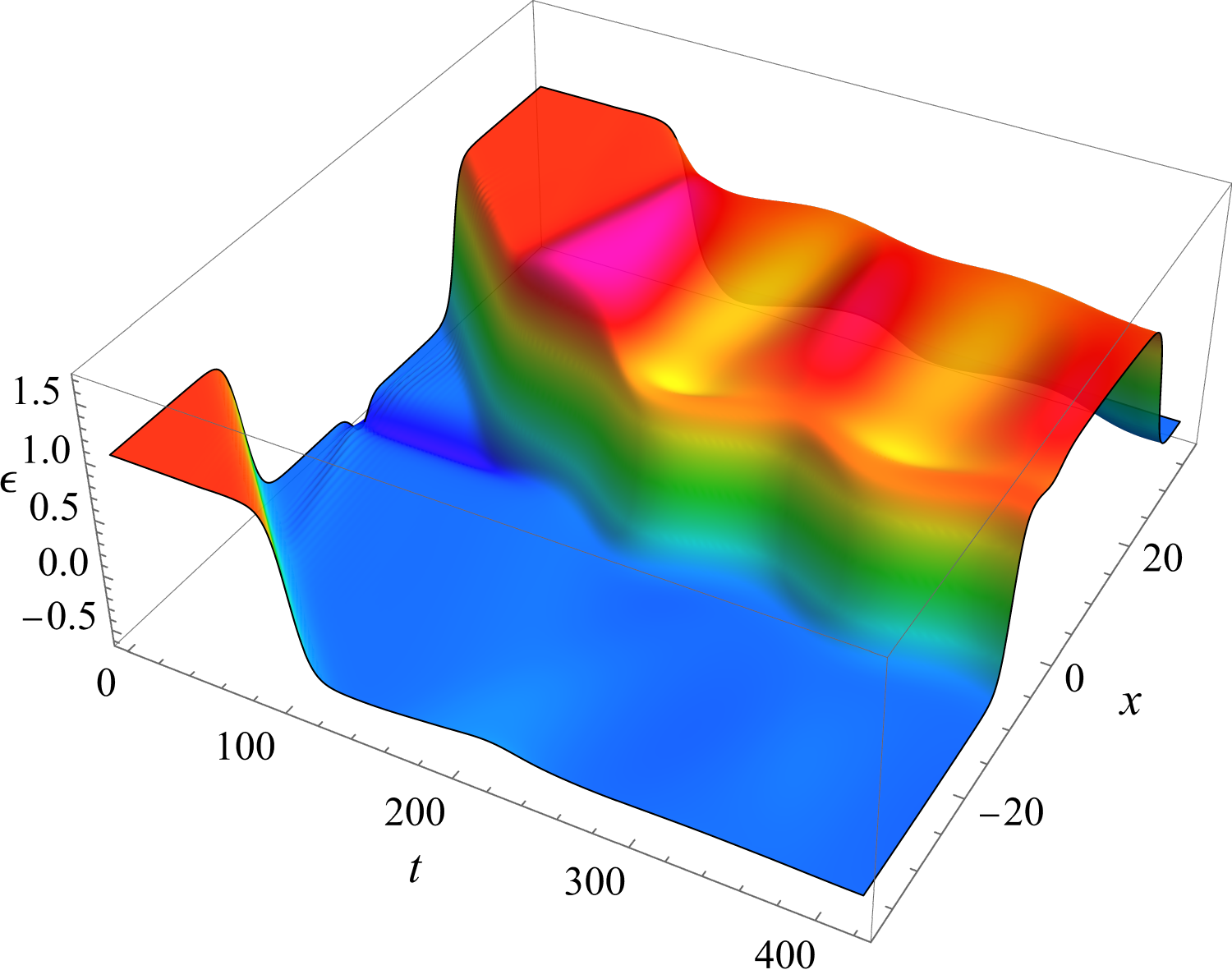}\label{fig:21}}
		\subfigure[]{\includegraphics[width=.49\linewidth]{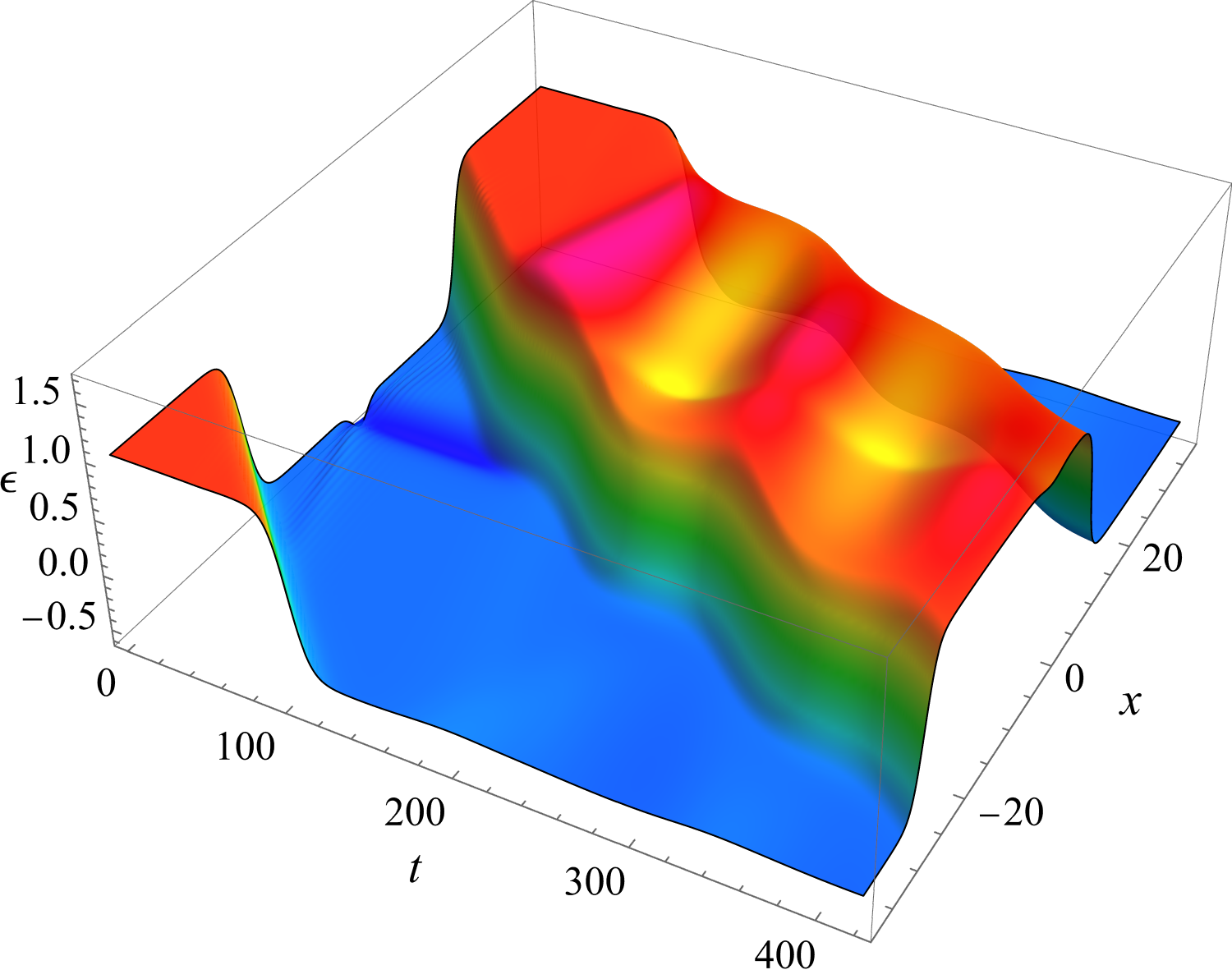}\label{fig:22}}
		\subfigure[]{\includegraphics[width=.49\linewidth]{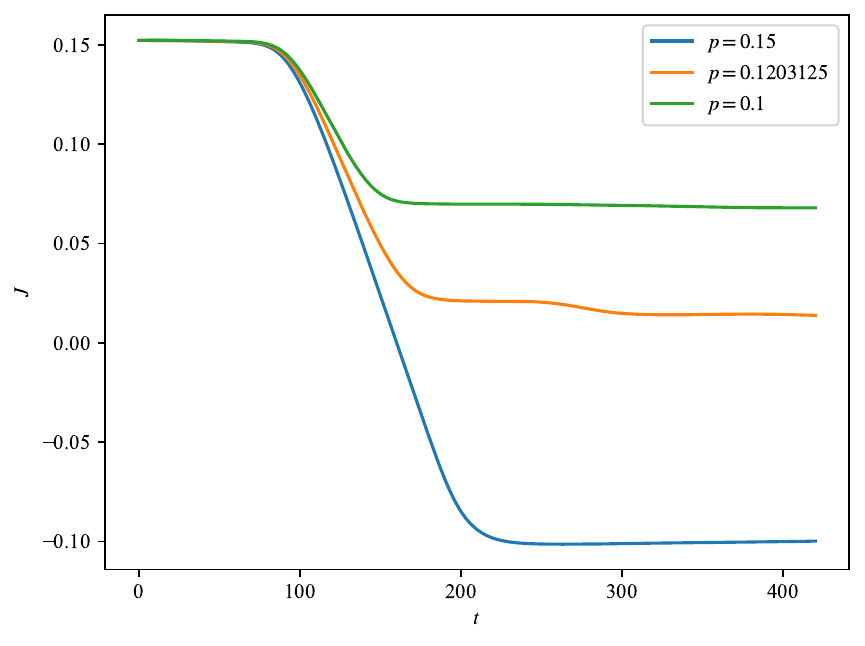}\label{fig:23}}
		\caption{Real-time dynamics of the energy density of the moving domain at flow velocity $v=0.18$, under the action of inhomogeneous scalar sources (\ref{eq:4.1}) with amplitudes $p=0.15$ (a), $p=0.1203125$ (b) and $p=0.1$ (c). The initial phase-separated state as background is generated from the spinodal decomposition of steady flow with flow velocity $v=0.2$ and energy density $\epsilon=0.325$. (d) The average momentum $J$ as a function of time. Curves of different colors represent situations with different amplitudes.}
		\label{fig:20-23}
	\end{center}
\end{figure}

\begin{figure}
	\begin{center}
		\subfigure[]{\includegraphics[width=.49\linewidth]{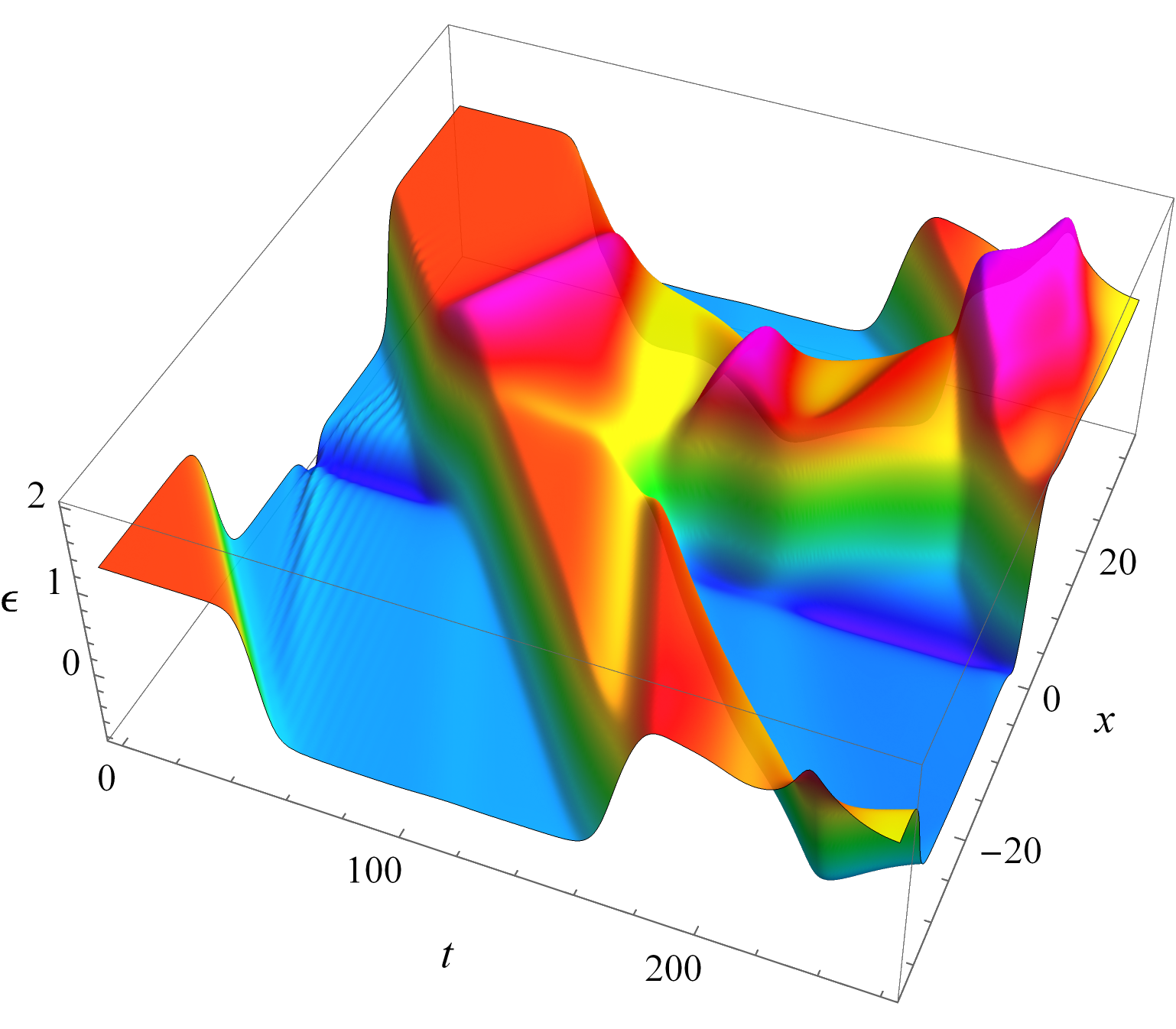}\label{fig:24}}
		\subfigure[]{\includegraphics[width=.49\linewidth]{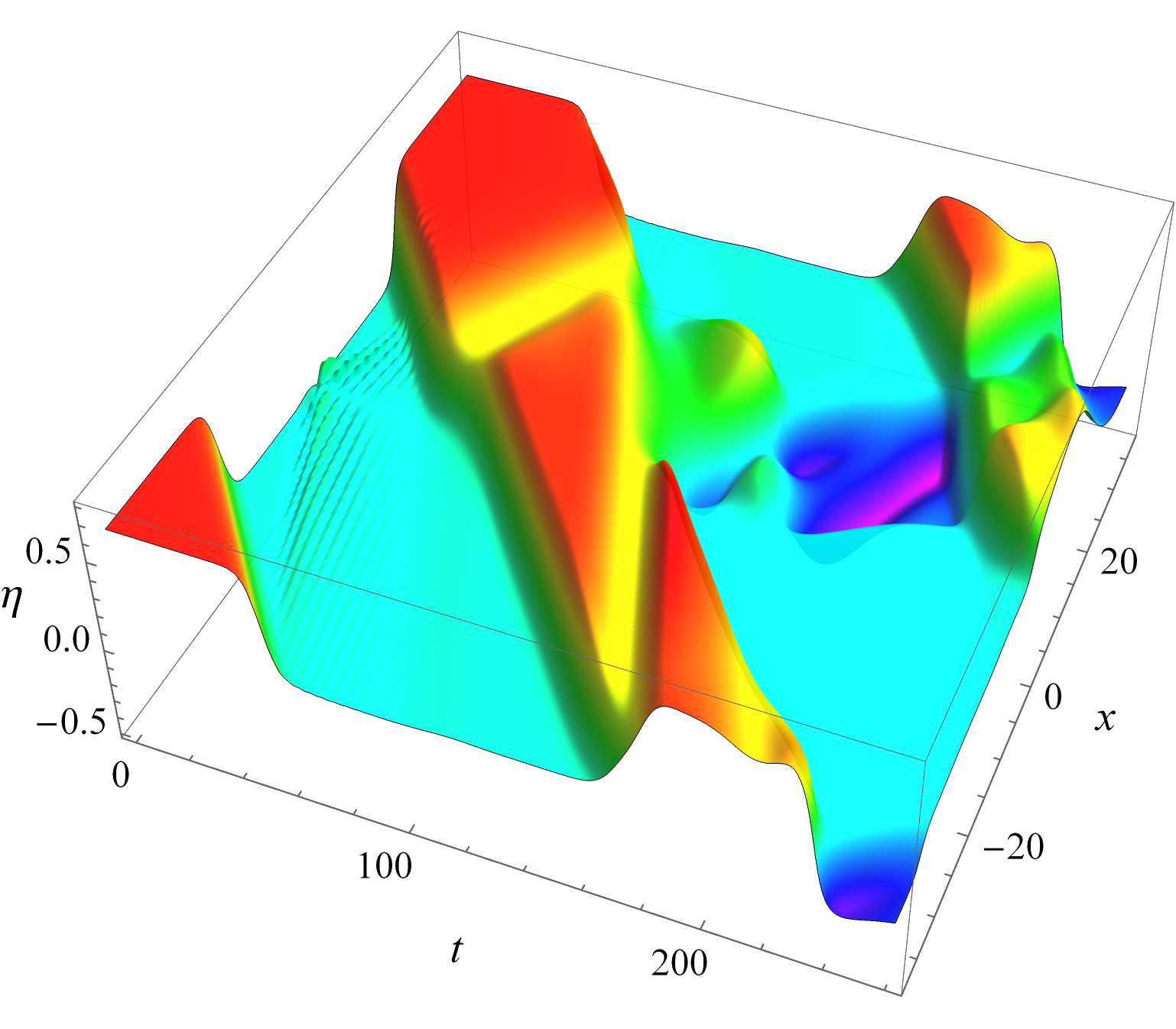}\label{fig:25}}
		\caption{Real-time dynamics of the energy density (a) and the momentum density (b) of the moving domain at flow velocity $v=0.37$, under the action of an inhomogeneous scalar source (\ref{eq:4.1}) with amplitude $p=0.2$. The initial phase-separated state as background is generated from the spinodal decomposition of steady flow with flow velocity $v=0.4$ and energy density $\epsilon=0.434$.}
		\label{fig:24-25}
	\end{center}
\end{figure}

For the situation of low flow velocity, the real-time dynamics of the energy density of the system are shown in figure \ref{fig:20-23}, from which one can clearly observe three characteristic dynamical behaviors of the moving domain when interacting with fixed obstacles of different strengths.
As shown in figure \ref{fig:20}, after hitting an obstacle with a high potential barrier, the moving domain will be rebounded. 
In the process, the momentum of the system decreases and its sign also changes.
It is worth noting that an energy wave will not be bounced off, but will pass through the energy well and continue forward, as shown in the third stage of figure \ref{fig:14}.
Such results indicate that the obstacle has a stronger blocking effect on the energy gap connected by the domain wall than on energy fluctuation.
There is also sufficient reason to infer that the larger the energy gap, the greater the obstruction effect.
As the amplitude $p$ of the scalar source decreases, there is a critical amplitude at which the moving domain becomes trapped by the obstacle, as shown in figure \ref{fig:21}.
At the moment of collision, the momentum of the system rapidly decays to near zero, as shown in figure \ref{fig:23}.
Such pinning behavior depends on the balance between the blocking effect of the obstacle, characterized by the amplitude $p$, and the impingement effect of the moving domain, characterized by the velocity $v$.
Amplitudes smaller than this threshold will make the obstacle insufficient to block the motion of the moving domain, causing the moving domain to completely pass through the obstacle, as shown in figure \ref{fig:22}. 
However, due to the attenuation of the momentum of the system, the velocity of the moving domain will gradually decrease, resulting in rebound behavior at the next interaction.
Actually, for a scalar source with a fixed small amplitude, there is a critical velocity that causes the moving domain to be imprisoned by the obstacle, and velocities less (greater) than this critical value will cause rebound (pass-through) behavior.

On the other hand, for the situation of high flow velocity, rather than a critical point corresponding to pinning behavior, there is a continuous interval in the amplitude that separates rebound and pass-through behaviors. 
For amplitudes in such an intermediate range, the moving domain will split during the process of passing the obstacle, generating two high-energy phases moving in opposite directions, as shown in figure \ref{fig:24-25}.
Among them, the one that passes through the obstacle carries positive momentum, while the other one that is bounced off carries negative momentum.
The spatial ratio of the two components depends on the amplitude of the scalar source.
There is an amplitude within the interval that leads to continuous splitting behavior, at which the total momentum of the system is zero.
The two ends of this interval correspond to the space proportion of one of the components being zero.
Over time, these two components will eventually collide together to form a moving domain with a new velocity that determines the next dynamical behavior.
Regardless of which of the above dynamical processes the system undergoes to evolve, the final state is a static phase-separated configuration with zero momentum.

\section{Conclusion}\label{sec:C}
In this paper, we systematically investigate the thermodynamic properties and nonequilibrium dynamics of steady flows in a holographic first-order phase transition model.
In particular, we simulate a local obstacle by introducing an inhomogeneous scalar external source, revealing the dynamical behavior of the moving system when interacting with a fixed obstacle.

For the thermodynamic properties of the equilibrium state, similar to the case of a static fluid, the phase structure of the free energy density of the steady flow also exhibits a swallowtail shape, indicating a thermal first-order phase transition.
The relationship between the characteristic physical parameters describing the first-order phase transition, phase transition temperature and latent heat, and the flow velocity is revealed.
With the increase of flow velocity, the phase transition temperature decreases and latent heat increases, indicating a gradually increasing potential barrier.
In particular, for the steady flow, significantly different from the static fluid, there is a gap each in the momentum density and longitudinal pressure between the coexisting phases at the phase transition temperature.
Moreover, such gaps can be quantitatively characterized by flow velocity and latent heat.

In addition, the steady flow also exhibits consistency in terms of thermodynamic stability and dynamical stability.
On the one hand, a thermodynamically unstable steady flow with a negative specific heat suffers from linear dynamical instability.
Under an arbitrarily small perturbation, a dynamical transition will occur, leading to the emergence of a phase-separated configuration with non-zero momentum.
For such a time-dependent inhomogeneous {stationary state}, there exists a Killing vector $\partial_{t}-v\partial_{x}$ with a uniform flow velocity $v$.
On the other hand, a thermodynamically metastable steady flow is dynamical unstable at the nonlinear level.
The occurrence of dynamical transition requires that the intensity of the disturbance exceeds a threshold to push the system across a critical state that acts as a dynamical barrier and then evolve to another local ground state, a moving phase-separated state.
For the thermodynamically stable steady flow, the system is dynamically stable and located in a global ground state.

Finally and most importantly, by introducing an inhomogeneous scalar external source, the real-time dynamics of a moving holographic first-order phase transition system under the action of a fixed obstacle are revealed.
Since the local structure of the scalar source invalidates the Killing vector $\partial_{t}-v\partial_{x}$ with non-zero velocity, the system keeps losing momentum during the dynamical process until it degenerates to a static state with a larger entropy and a Killing vector $\partial_{t}$.
Such results show that in a closed system with fixed energy, there is a dynamical path that allows the system to evolve from a moving {stationary state} to a static state, indicating the higher stability of the static state.
In particular, due to the blocking effect of the obstacle on motion, the moving high-energy phase exhibits four characteristic dynamical behaviors when interacting with it, rebounding, pinning, passing through, and splitting, depending on the competitive relationship between the flow velocity of the moving domain and the strength of the obstacle.
Such results provide an intuitive physical picture for experiments with strongly coupled fluids.

The interaction between moving objects and obstacles is an interesting topic, especially for the domain walls which are crucial to the dynamical transport properties of the system.
In our work, the domain wall of the effectively one-dimensional system is only a point, which makes the dynamical properties of the domain wall neglected during the interaction process.
In the future, an important extension is to consider effectively two-dimensional or even three-dimensional transport systems, corresponding to linear and planar domain walls respectively.
In that case, the domain wall will deform when interacting with an obstacle, further revealing the effects of surface tension and distortion energy on the dynamical properties of the transport system.
In addition, it is worthwhile to further explore the dynamical properties of other strongly coupled transport systems, such as holographic superfluids and turbulence, under interaction with obstacles.
{On the other hand, the boundary effects ignored in our work are also an interesting topic. 
Through different boundary conditions, we can simulate the interaction between the physical system and the external environment. 
At this time, the physical quantities of the system may not be conserved, such as energy and momentum, which is different from the closed system in this work. 
For an open boundary, the physical phenomena may change significantly, for example, the energy and momentum of a steady flow within a fixed interval may be continuously lost. 
Such a result will lead to a continuous weakening of the interaction between the steady flow and the obstacle, thereby changing the dynamic behavior of the system.}
Such studies will definitely provide even richer physical phenomena.

\appendix

\section*{Acknowledgement}
This work is supported in part by the National Natural Science Foundation of China under Grant Nos. {12035016}, 12075026, 12275350, {12361141825}, {12375058} and 12447129 as well as 
the National Key Research and Development Program of China with Grant No. 2021YFC2203001, and the China Postdoctoral Science Foundation with Grant No. 2024M760691.

\bibliography{references}

\begin{thebibliography}{61}%
\makeatletter
\providecommand \@ifxundefined [1]{%
 \@ifx{#1\undefined}
}%
\providecommand \@ifnum [1]{%
 \ifnum #1\expandafter \@firstoftwo
 \else \expandafter \@secondoftwo
 \fi
}%
\providecommand \@ifx [1]{%
 \ifx #1\expandafter \@firstoftwo
 \else \expandafter \@secondoftwo
 \fi
}%
\providecommand \natexlab [1]{#1}%
\providecommand \enquote  [1]{``#1''}%
\providecommand \bibnamefont  [1]{#1}%
\providecommand \bibfnamefont [1]{#1}%
\providecommand \citenamefont [1]{#1}%
\providecommand \href@noop [0]{\@secondoftwo}%
\providecommand \href [0]{\begingroup \@sanitize@url \@href}%
\providecommand \@href[1]{\@@startlink{#1}\@@href}%
\providecommand \@@href[1]{\endgroup#1\@@endlink}%
\providecommand \@sanitize@url [0]{\catcode `\\12\catcode `\$12\catcode
  `\&12\catcode `\#12\catcode `\^12\catcode `\_12\catcode `\%12\relax}%
\providecommand \@@startlink[1]{}%
\providecommand \@@endlink[0]{}%
\providecommand \url  [0]{\begingroup\@sanitize@url \@url }%
\providecommand \@url [1]{\endgroup\@href {#1}{\urlprefix }}%
\providecommand \urlprefix  [0]{URL }%
\providecommand \Eprint [0]{\href }%
\providecommand \doibase [0]{https://doi.org/}%
\providecommand \selectlanguage [0]{\@gobble}%
\providecommand \bibinfo  [0]{\@secondoftwo}%
\providecommand \bibfield  [0]{\@secondoftwo}%
\providecommand \translation [1]{[#1]}%
\providecommand \BibitemOpen [0]{}%
\providecommand \bibitemStop [0]{}%
\providecommand \bibitemNoStop [0]{.\EOS\space}%
\providecommand \EOS [0]{\spacefactor3000\relax}%
\providecommand \BibitemShut  [1]{\csname bibitem#1\endcsname}%
\let\auto@bib@innerbib\@empty
\bibitem [{\citenamefont {Sch\"afer}\ and\ \citenamefont
  {Teaney}(2009)}]{Schafer:2009dj}%
  \BibitemOpen
  \bibfield  {author} {\bibinfo {author} {\bibfnamefont {T.}~\bibnamefont
  {Sch\"afer}}\ and\ \bibinfo {author} {\bibfnamefont {D.}~\bibnamefont
  {Teaney}},\ }\bibfield  {title} {\bibinfo {title} {{Nearly Perfect Fluidity:
  From Cold Atomic Gases to Hot Quark Gluon Plasmas}},\ }\href
  {https://doi.org/10.1088/0034-4885/72/12/126001} {\bibfield  {journal}
  {\bibinfo  {journal} {Rept. Prog. Phys.}\ }\textbf {\bibinfo {volume} {72}},\
  \bibinfo {pages} {126001} (\bibinfo {year} {2009})},\ \Eprint
  {https://arxiv.org/abs/0904.3107} {arXiv:0904.3107 [hep-ph]} \BibitemShut
  {NoStop}%
\bibitem [{\citenamefont {Alford}\ \emph {et~al.}(2018)\citenamefont {Alford},
  \citenamefont {Bovard}, \citenamefont {Hanauske}, \citenamefont {Rezzolla},\
  and\ \citenamefont {Schwenzer}}]{Alford:2017rxf}%
  \BibitemOpen
  \bibfield  {author} {\bibinfo {author} {\bibfnamefont {M.~G.}\ \bibnamefont
  {Alford}}, \bibinfo {author} {\bibfnamefont {L.}~\bibnamefont {Bovard}},
  \bibinfo {author} {\bibfnamefont {M.}~\bibnamefont {Hanauske}}, \bibinfo
  {author} {\bibfnamefont {L.}~\bibnamefont {Rezzolla}},\ and\ \bibinfo
  {author} {\bibfnamefont {K.}~\bibnamefont {Schwenzer}},\ }\bibfield  {title}
  {\bibinfo {title} {{Viscous Dissipation and Heat Conduction in Binary
  Neutron-Star Mergers}},\ }\href
  {https://doi.org/10.1103/PhysRevLett.120.041101} {\bibfield  {journal}
  {\bibinfo  {journal} {Phys. Rev. Lett.}\ }\textbf {\bibinfo {volume} {120}},\
  \bibinfo {pages} {041101} (\bibinfo {year} {2018})},\ \Eprint
  {https://arxiv.org/abs/1707.09475} {arXiv:1707.09475 [gr-qc]} \BibitemShut
  {NoStop}%
\bibitem [{\citenamefont {Maldacena}(1998)}]{Maldacena:1997re}%
  \BibitemOpen
  \bibfield  {author} {\bibinfo {author} {\bibfnamefont {J.~M.}\ \bibnamefont
  {Maldacena}},\ }\bibfield  {title} {\bibinfo {title} {{The Large N limit of
  superconformal field theories and supergravity}},\ }\href
  {https://doi.org/10.1023/A:1026654312961} {\bibfield  {journal} {\bibinfo
  {journal} {Adv. Theor. Math. Phys.}\ }\textbf {\bibinfo {volume} {2}},\
  \bibinfo {pages} {231} (\bibinfo {year} {1998})},\ \Eprint
  {https://arxiv.org/abs/hep-th/9711200} {arXiv:hep-th/9711200} \BibitemShut
  {NoStop}%
\bibitem [{\citenamefont {Gubser}\ \emph {et~al.}(1998)\citenamefont {Gubser},
  \citenamefont {Klebanov},\ and\ \citenamefont {Polyakov}}]{Gubser:1998bc}%
  \BibitemOpen
  \bibfield  {author} {\bibinfo {author} {\bibfnamefont {S.~S.}\ \bibnamefont
  {Gubser}}, \bibinfo {author} {\bibfnamefont {I.~R.}\ \bibnamefont
  {Klebanov}},\ and\ \bibinfo {author} {\bibfnamefont {A.~M.}\ \bibnamefont
  {Polyakov}},\ }\bibfield  {title} {\bibinfo {title} {{Gauge theory
  correlators from noncritical string theory}},\ }\href
  {https://doi.org/10.1016/S0370-2693(98)00377-3} {\bibfield  {journal}
  {\bibinfo  {journal} {Phys. Lett. B}\ }\textbf {\bibinfo {volume} {428}},\
  \bibinfo {pages} {105} (\bibinfo {year} {1998})},\ \Eprint
  {https://arxiv.org/abs/hep-th/9802109} {arXiv:hep-th/9802109} \BibitemShut
  {NoStop}%
\bibitem [{\citenamefont {Witten}(1998{\natexlab{a}})}]{Witten:1998qj}%
  \BibitemOpen
  \bibfield  {author} {\bibinfo {author} {\bibfnamefont {E.}~\bibnamefont
  {Witten}},\ }\bibfield  {title} {\bibinfo {title} {{Anti-de Sitter space and
  holography}},\ }\href {https://doi.org/10.4310/ATMP.1998.v2.n2.a2} {\bibfield
   {journal} {\bibinfo  {journal} {Adv. Theor. Math. Phys.}\ }\textbf {\bibinfo
  {volume} {2}},\ \bibinfo {pages} {253} (\bibinfo {year}
  {1998}{\natexlab{a}})},\ \Eprint {https://arxiv.org/abs/hep-th/9802150}
  {arXiv:hep-th/9802150} \BibitemShut {NoStop}%
\bibitem [{\citenamefont {Witten}(1998{\natexlab{b}})}]{Witten:1998zw}%
  \BibitemOpen
  \bibfield  {author} {\bibinfo {author} {\bibfnamefont {E.}~\bibnamefont
  {Witten}},\ }\bibfield  {title} {\bibinfo {title} {{Anti-de Sitter space,
  thermal phase transition, and confinement in gauge theories}},\ }\href
  {https://doi.org/10.4310/ATMP.1998.v2.n3.a3} {\bibfield  {journal} {\bibinfo
  {journal} {Adv. Theor. Math. Phys.}\ }\textbf {\bibinfo {volume} {2}},\
  \bibinfo {pages} {505} (\bibinfo {year} {1998}{\natexlab{b}})},\ \Eprint
  {https://arxiv.org/abs/hep-th/9803131} {arXiv:hep-th/9803131} \BibitemShut
  {NoStop}%
\bibitem [{\citenamefont {Chesler}\ and\ \citenamefont
  {Yaffe}(2014)}]{Chesler:2013lia}%
  \BibitemOpen
  \bibfield  {author} {\bibinfo {author} {\bibfnamefont {P.~M.}\ \bibnamefont
  {Chesler}}\ and\ \bibinfo {author} {\bibfnamefont {L.~G.}\ \bibnamefont
  {Yaffe}},\ }\bibfield  {title} {\bibinfo {title} {{Numerical solution of
  gravitational dynamics in asymptotically anti-de Sitter spacetimes}},\ }\href
  {https://doi.org/10.1007/JHEP07(2014)086} {\bibfield  {journal} {\bibinfo
  {journal} {JHEP}\ }\textbf {\bibinfo {volume} {07}},\ \bibinfo {pages}
  {086}},\ \Eprint {https://arxiv.org/abs/1309.1439} {arXiv:1309.1439 [hep-th]}
  \BibitemShut {NoStop}%
\bibitem [{\citenamefont {Gubser}\ \emph {et~al.}(2008)\citenamefont {Gubser},
  \citenamefont {Nellore}, \citenamefont {Pufu},\ and\ \citenamefont
  {Rocha}}]{Gubser:2008yx}%
  \BibitemOpen
  \bibfield  {author} {\bibinfo {author} {\bibfnamefont {S.~S.}\ \bibnamefont
  {Gubser}}, \bibinfo {author} {\bibfnamefont {A.}~\bibnamefont {Nellore}},
  \bibinfo {author} {\bibfnamefont {S.~S.}\ \bibnamefont {Pufu}},\ and\
  \bibinfo {author} {\bibfnamefont {F.~D.}\ \bibnamefont {Rocha}},\ }\bibfield
  {title} {\bibinfo {title} {{Thermodynamics and bulk viscosity of approximate
  black hole duals to finite temperature quantum chromodynamics}},\ }\href
  {https://doi.org/10.1103/PhysRevLett.101.131601} {\bibfield  {journal}
  {\bibinfo  {journal} {Phys. Rev. Lett.}\ }\textbf {\bibinfo {volume} {101}},\
  \bibinfo {pages} {131601} (\bibinfo {year} {2008})},\ \Eprint
  {https://arxiv.org/abs/0804.1950} {arXiv:0804.1950 [hep-th]} \BibitemShut
  {NoStop}%
\bibitem [{\citenamefont {Gubser}\ and\ \citenamefont
  {Nellore}(2008)}]{Gubser:2008ny}%
  \BibitemOpen
  \bibfield  {author} {\bibinfo {author} {\bibfnamefont {S.~S.}\ \bibnamefont
  {Gubser}}\ and\ \bibinfo {author} {\bibfnamefont {A.}~\bibnamefont
  {Nellore}},\ }\bibfield  {title} {\bibinfo {title} {{Mimicking the QCD
  equation of state with a dual black hole}},\ }\href
  {https://doi.org/10.1103/PhysRevD.78.086007} {\bibfield  {journal} {\bibinfo
  {journal} {Phys. Rev. D}\ }\textbf {\bibinfo {volume} {78}},\ \bibinfo
  {pages} {086007} (\bibinfo {year} {2008})},\ \Eprint
  {https://arxiv.org/abs/0804.0434} {arXiv:0804.0434 [hep-th]} \BibitemShut
  {NoStop}%
\bibitem [{\citenamefont {Janik}\ \emph
  {et~al.}(2016{\natexlab{a}})\citenamefont {Janik}, \citenamefont
  {Jankowski},\ and\ \citenamefont {Soltanpanahi}}]{Janik:2015iry}%
  \BibitemOpen
  \bibfield  {author} {\bibinfo {author} {\bibfnamefont {R.~A.}\ \bibnamefont
  {Janik}}, \bibinfo {author} {\bibfnamefont {J.}~\bibnamefont {Jankowski}},\
  and\ \bibinfo {author} {\bibfnamefont {H.}~\bibnamefont {Soltanpanahi}},\
  }\bibfield  {title} {\bibinfo {title} {{Nonequilibrium Dynamics and Phase
  Transitions in Holographic Models}},\ }\href
  {https://doi.org/10.1103/PhysRevLett.117.091603} {\bibfield  {journal}
  {\bibinfo  {journal} {Phys. Rev. Lett.}\ }\textbf {\bibinfo {volume} {117}},\
  \bibinfo {pages} {091603} (\bibinfo {year} {2016}{\natexlab{a}})},\ \Eprint
  {https://arxiv.org/abs/1512.06871} {arXiv:1512.06871 [hep-th]} \BibitemShut
  {NoStop}%
\bibitem [{\citenamefont {Janik}\ \emph
  {et~al.}(2016{\natexlab{b}})\citenamefont {Janik}, \citenamefont
  {Jankowski},\ and\ \citenamefont {Soltanpanahi}}]{Janik:2016btb}%
  \BibitemOpen
  \bibfield  {author} {\bibinfo {author} {\bibfnamefont {R.~A.}\ \bibnamefont
  {Janik}}, \bibinfo {author} {\bibfnamefont {J.}~\bibnamefont {Jankowski}},\
  and\ \bibinfo {author} {\bibfnamefont {H.}~\bibnamefont {Soltanpanahi}},\
  }\bibfield  {title} {\bibinfo {title} {{Quasinormal modes and the phase
  structure of strongly coupled matter}},\ }\href
  {https://doi.org/10.1007/JHEP06(2016)047} {\bibfield  {journal} {\bibinfo
  {journal} {JHEP}\ }\textbf {\bibinfo {volume} {06}},\ \bibinfo {pages}
  {047}},\ \Eprint {https://arxiv.org/abs/1603.05950} {arXiv:1603.05950
  [hep-th]} \BibitemShut {NoStop}%
\bibitem [{\citenamefont {Gregory}\ and\ \citenamefont
  {Laflamme}(1993)}]{Gregory:1993vy}%
  \BibitemOpen
  \bibfield  {author} {\bibinfo {author} {\bibfnamefont {R.}~\bibnamefont
  {Gregory}}\ and\ \bibinfo {author} {\bibfnamefont {R.}~\bibnamefont
  {Laflamme}},\ }\bibfield  {title} {\bibinfo {title} {{Black strings and
  p-branes are unstable}},\ }\href
  {https://doi.org/10.1103/PhysRevLett.70.2837} {\bibfield  {journal} {\bibinfo
   {journal} {Phys. Rev. Lett.}\ }\textbf {\bibinfo {volume} {70}},\ \bibinfo
  {pages} {2837} (\bibinfo {year} {1993})},\ \Eprint
  {https://arxiv.org/abs/hep-th/9301052} {arXiv:hep-th/9301052} \BibitemShut
  {NoStop}%
\bibitem [{\citenamefont {Gregory}\ and\ \citenamefont
  {Laflamme}(1994)}]{Gregory:1994bj}%
  \BibitemOpen
  \bibfield  {author} {\bibinfo {author} {\bibfnamefont {R.}~\bibnamefont
  {Gregory}}\ and\ \bibinfo {author} {\bibfnamefont {R.}~\bibnamefont
  {Laflamme}},\ }\bibfield  {title} {\bibinfo {title} {{The Instability of
  charged black strings and p-branes}},\ }\href
  {https://doi.org/10.1016/0550-3213(94)90206-2} {\bibfield  {journal}
  {\bibinfo  {journal} {Nucl. Phys. B}\ }\textbf {\bibinfo {volume} {428}},\
  \bibinfo {pages} {399} (\bibinfo {year} {1994})},\ \Eprint
  {https://arxiv.org/abs/hep-th/9404071} {arXiv:hep-th/9404071} \BibitemShut
  {NoStop}%
\bibitem [{\citenamefont {Attems}\ \emph
  {et~al.}(2017{\natexlab{a}})\citenamefont {Attems}, \citenamefont {Bea},
  \citenamefont {Casalderrey-Solana}, \citenamefont {Mateos}, \citenamefont
  {Triana},\ and\ \citenamefont {Zilhao}}]{Attems:2017ezz}%
  \BibitemOpen
  \bibfield  {author} {\bibinfo {author} {\bibfnamefont {M.}~\bibnamefont
  {Attems}}, \bibinfo {author} {\bibfnamefont {Y.}~\bibnamefont {Bea}},
  \bibinfo {author} {\bibfnamefont {J.}~\bibnamefont {Casalderrey-Solana}},
  \bibinfo {author} {\bibfnamefont {D.}~\bibnamefont {Mateos}}, \bibinfo
  {author} {\bibfnamefont {M.}~\bibnamefont {Triana}},\ and\ \bibinfo {author}
  {\bibfnamefont {M.}~\bibnamefont {Zilhao}},\ }\bibfield  {title} {\bibinfo
  {title} {{Phase Transitions, Inhomogeneous Horizons and Second-Order
  Hydrodynamics}},\ }\href {https://doi.org/10.1007/JHEP06(2017)129} {\bibfield
   {journal} {\bibinfo  {journal} {JHEP}\ }\textbf {\bibinfo {volume} {06}},\
  \bibinfo {pages} {129}},\ \Eprint {https://arxiv.org/abs/1703.02948}
  {arXiv:1703.02948 [hep-th]} \BibitemShut {NoStop}%
\bibitem [{\citenamefont {Janik}\ \emph {et~al.}(2017)\citenamefont {Janik},
  \citenamefont {Jankowski},\ and\ \citenamefont
  {Soltanpanahi}}]{Janik:2017ykj}%
  \BibitemOpen
  \bibfield  {author} {\bibinfo {author} {\bibfnamefont {R.~A.}\ \bibnamefont
  {Janik}}, \bibinfo {author} {\bibfnamefont {J.}~\bibnamefont {Jankowski}},\
  and\ \bibinfo {author} {\bibfnamefont {H.}~\bibnamefont {Soltanpanahi}},\
  }\bibfield  {title} {\bibinfo {title} {{Real-Time dynamics and phase
  separation in a holographic first order phase transition}},\ }\href
  {https://doi.org/10.1103/PhysRevLett.119.261601} {\bibfield  {journal}
  {\bibinfo  {journal} {Phys. Rev. Lett.}\ }\textbf {\bibinfo {volume} {119}},\
  \bibinfo {pages} {261601} (\bibinfo {year} {2017})},\ \Eprint
  {https://arxiv.org/abs/1704.05387} {arXiv:1704.05387 [hep-th]} \BibitemShut
  {NoStop}%
\bibitem [{\citenamefont {Bellantuono}\ \emph {et~al.}(2019)\citenamefont
  {Bellantuono}, \citenamefont {Janik}, \citenamefont {Jankowski},\ and\
  \citenamefont {Soltanpanahi}}]{Bellantuono:2019wbn}%
  \BibitemOpen
  \bibfield  {author} {\bibinfo {author} {\bibfnamefont {L.}~\bibnamefont
  {Bellantuono}}, \bibinfo {author} {\bibfnamefont {R.~A.}\ \bibnamefont
  {Janik}}, \bibinfo {author} {\bibfnamefont {J.}~\bibnamefont {Jankowski}},\
  and\ \bibinfo {author} {\bibfnamefont {H.}~\bibnamefont {Soltanpanahi}},\
  }\bibfield  {title} {\bibinfo {title} {{Dynamics near a first order phase
  transition}},\ }\href {https://doi.org/10.1007/JHEP10(2019)146} {\bibfield
  {journal} {\bibinfo  {journal} {JHEP}\ }\textbf {\bibinfo {volume} {10}},\
  \bibinfo {pages} {146}},\ \Eprint {https://arxiv.org/abs/1906.00061}
  {arXiv:1906.00061 [hep-th]} \BibitemShut {NoStop}%
\bibitem [{\citenamefont {Attems}\ \emph {et~al.}(2020)\citenamefont {Attems},
  \citenamefont {Bea}, \citenamefont {Casalderrey-Solana}, \citenamefont
  {Mateos},\ and\ \citenamefont {Zilh\~ao}}]{Attems:2019yqn}%
  \BibitemOpen
  \bibfield  {author} {\bibinfo {author} {\bibfnamefont {M.}~\bibnamefont
  {Attems}}, \bibinfo {author} {\bibfnamefont {Y.}~\bibnamefont {Bea}},
  \bibinfo {author} {\bibfnamefont {J.}~\bibnamefont {Casalderrey-Solana}},
  \bibinfo {author} {\bibfnamefont {D.}~\bibnamefont {Mateos}},\ and\ \bibinfo
  {author} {\bibfnamefont {M.}~\bibnamefont {Zilh\~ao}},\ }\bibfield  {title}
  {\bibinfo {title} {{Dynamics of Phase Separation from Holography}},\ }\href
  {https://doi.org/10.1007/JHEP01(2020)106} {\bibfield  {journal} {\bibinfo
  {journal} {JHEP}\ }\textbf {\bibinfo {volume} {01}},\ \bibinfo {pages}
  {106}},\ \Eprint {https://arxiv.org/abs/1905.12544} {arXiv:1905.12544
  [hep-th]} \BibitemShut {NoStop}%
\bibitem [{\citenamefont {Attems}(2021)}]{Attems:2020qkg}%
  \BibitemOpen
  \bibfield  {author} {\bibinfo {author} {\bibfnamefont {M.}~\bibnamefont
  {Attems}},\ }\bibfield  {title} {\bibinfo {title} {{Holographic approach of
  the spinodal instability to criticality}},\ }\href
  {https://doi.org/10.1007/JHEP08(2021)155} {\bibfield  {journal} {\bibinfo
  {journal} {JHEP}\ }\textbf {\bibinfo {volume} {08}},\ \bibinfo {pages}
  {155}},\ \Eprint {https://arxiv.org/abs/2012.15687} {arXiv:2012.15687
  [hep-th]} \BibitemShut {NoStop}%
\bibitem [{\citenamefont {Caddeo}\ \emph {et~al.}(2024)\citenamefont {Caddeo},
  \citenamefont {Henriksson}, \citenamefont {Hoyos},\ and\ \citenamefont
  {Sanchez-Garitaonandia}}]{Caddeo:2024lfk}%
  \BibitemOpen
  \bibfield  {author} {\bibinfo {author} {\bibfnamefont {A.}~\bibnamefont
  {Caddeo}}, \bibinfo {author} {\bibfnamefont {O.}~\bibnamefont {Henriksson}},
  \bibinfo {author} {\bibfnamefont {C.}~\bibnamefont {Hoyos}},\ and\ \bibinfo
  {author} {\bibfnamefont {M.}~\bibnamefont {Sanchez-Garitaonandia}},\
  }\bibfield  {title} {\bibinfo {title} {{Spinodal slowing down and scaling in
  a holographic model}},\ }\href {https://doi.org/10.1007/JHEP08(2024)091}
  {\bibfield  {journal} {\bibinfo  {journal} {JHEP}\ }\textbf {\bibinfo
  {volume} {08}},\ \bibinfo {pages} {091}},\ \Eprint
  {https://arxiv.org/abs/2406.15297} {arXiv:2406.15297 [hep-th]} \BibitemShut
  {NoStop}%
\bibitem [{\citenamefont {Bea}\ \emph {et~al.}(2021{\natexlab{a}})\citenamefont
  {Bea}, \citenamefont {Dias}, \citenamefont {Giannakopoulos}, \citenamefont
  {Mateos}, \citenamefont {Sanchez-Garitaonandia}, \citenamefont {Santos},\
  and\ \citenamefont {Zilh\~ao}}]{Bea:2020ees}%
  \BibitemOpen
  \bibfield  {author} {\bibinfo {author} {\bibfnamefont {Y.}~\bibnamefont
  {Bea}}, \bibinfo {author} {\bibfnamefont {O.~J.~C.}\ \bibnamefont {Dias}},
  \bibinfo {author} {\bibfnamefont {T.}~\bibnamefont {Giannakopoulos}},
  \bibinfo {author} {\bibfnamefont {D.}~\bibnamefont {Mateos}}, \bibinfo
  {author} {\bibfnamefont {M.}~\bibnamefont {Sanchez-Garitaonandia}}, \bibinfo
  {author} {\bibfnamefont {J.~E.}\ \bibnamefont {Santos}},\ and\ \bibinfo
  {author} {\bibfnamefont {M.}~\bibnamefont {Zilh\~ao}},\ }\bibfield  {title}
  {\bibinfo {title} {{Crossing a large-$N$ phase transition at finite
  volume}},\ }\href {https://doi.org/10.1007/JHEP02(2021)061} {\bibfield
  {journal} {\bibinfo  {journal} {JHEP}\ }\textbf {\bibinfo {volume} {02}},\
  \bibinfo {pages} {061}},\ \Eprint {https://arxiv.org/abs/2007.06467}
  {arXiv:2007.06467 [hep-th]} \BibitemShut {NoStop}%
\bibitem [{\citenamefont {Bea}\ \emph {et~al.}(2022{\natexlab{a}})\citenamefont
  {Bea}, \citenamefont {Casalderrey-Solana}, \citenamefont {Giannakopoulos},
  \citenamefont {Jansen}, \citenamefont {Mateos}, \citenamefont
  {Sanchez-Garitaonandia},\ and\ \citenamefont {Zilh\~ao}}]{Bea:2022mfb}%
  \BibitemOpen
  \bibfield  {author} {\bibinfo {author} {\bibfnamefont {Y.}~\bibnamefont
  {Bea}}, \bibinfo {author} {\bibfnamefont {J.}~\bibnamefont
  {Casalderrey-Solana}}, \bibinfo {author} {\bibfnamefont {T.}~\bibnamefont
  {Giannakopoulos}}, \bibinfo {author} {\bibfnamefont {A.}~\bibnamefont
  {Jansen}}, \bibinfo {author} {\bibfnamefont {D.}~\bibnamefont {Mateos}},
  \bibinfo {author} {\bibfnamefont {M.}~\bibnamefont {Sanchez-Garitaonandia}},\
  and\ \bibinfo {author} {\bibfnamefont {M.}~\bibnamefont {Zilh\~ao}},\
  }\bibfield  {title} {\bibinfo {title} {{Holographic bubbles with Jecco:
  expanding, collapsing and critical}},\ }\href
  {https://doi.org/10.1007/JHEP09(2022)008} {\bibfield  {journal} {\bibinfo
  {journal} {JHEP}\ }\textbf {\bibinfo {volume} {09}},\ \bibinfo {pages}
  {008}},\ \bibinfo {note} {[Erratum: JHEP 03, 225 (2023)]},\ \Eprint
  {https://arxiv.org/abs/2202.10503} {arXiv:2202.10503 [hep-th]} \BibitemShut
  {NoStop}%
\bibitem [{\citenamefont {Chen}\ \emph
  {et~al.}(2023{\natexlab{a}})\citenamefont {Chen}, \citenamefont {Liu},
  \citenamefont {Tian}, \citenamefont {Wang}, \citenamefont {Zhang},\ and\
  \citenamefont {Zhang}}]{Chen:2022cwi}%
  \BibitemOpen
  \bibfield  {author} {\bibinfo {author} {\bibfnamefont {Q.}~\bibnamefont
  {Chen}}, \bibinfo {author} {\bibfnamefont {Y.}~\bibnamefont {Liu}}, \bibinfo
  {author} {\bibfnamefont {Y.}~\bibnamefont {Tian}}, \bibinfo {author}
  {\bibfnamefont {B.}~\bibnamefont {Wang}}, \bibinfo {author} {\bibfnamefont
  {C.-Y.}\ \bibnamefont {Zhang}},\ and\ \bibinfo {author} {\bibfnamefont
  {H.}~\bibnamefont {Zhang}},\ }\bibfield  {title} {\bibinfo {title} {{Critical
  dynamics in holographic first-order phase transition}},\ }\href
  {https://doi.org/10.1007/JHEP01(2023)056} {\bibfield  {journal} {\bibinfo
  {journal} {JHEP}\ }\textbf {\bibinfo {volume} {01}},\ \bibinfo {pages}
  {056}},\ \Eprint {https://arxiv.org/abs/2209.12789} {arXiv:2209.12789
  [hep-th]} \BibitemShut {NoStop}%
\bibitem [{\citenamefont {Choptuik}\ \emph {et~al.}(1996)\citenamefont
  {Choptuik}, \citenamefont {Chmaj},\ and\ \citenamefont
  {Bizon}}]{Choptuik:1996yg}%
  \BibitemOpen
  \bibfield  {author} {\bibinfo {author} {\bibfnamefont {M.~W.}\ \bibnamefont
  {Choptuik}}, \bibinfo {author} {\bibfnamefont {T.}~\bibnamefont {Chmaj}},\
  and\ \bibinfo {author} {\bibfnamefont {P.}~\bibnamefont {Bizon}},\ }\bibfield
   {title} {\bibinfo {title} {{Critical behavior in gravitational collapse of a
  Yang-Mills field}},\ }\href {https://doi.org/10.1103/PhysRevLett.77.424}
  {\bibfield  {journal} {\bibinfo  {journal} {Phys. Rev. Lett.}\ }\textbf
  {\bibinfo {volume} {77}},\ \bibinfo {pages} {424} (\bibinfo {year} {1996})},\
  \Eprint {https://arxiv.org/abs/gr-qc/9603051} {arXiv:gr-qc/9603051}
  \BibitemShut {NoStop}%
\bibitem [{\citenamefont {Liebling}\ and\ \citenamefont
  {Choptuik}(1996)}]{Liebling:1996dx}%
  \BibitemOpen
  \bibfield  {author} {\bibinfo {author} {\bibfnamefont {S.~L.}\ \bibnamefont
  {Liebling}}\ and\ \bibinfo {author} {\bibfnamefont {M.~W.}\ \bibnamefont
  {Choptuik}},\ }\bibfield  {title} {\bibinfo {title} {{Black hole criticality
  in the Brans-Dicke model}},\ }\href
  {https://doi.org/10.1103/PhysRevLett.77.1424} {\bibfield  {journal} {\bibinfo
   {journal} {Phys. Rev. Lett.}\ }\textbf {\bibinfo {volume} {77}},\ \bibinfo
  {pages} {1424} (\bibinfo {year} {1996})},\ \Eprint
  {https://arxiv.org/abs/gr-qc/9606057} {arXiv:gr-qc/9606057} \BibitemShut
  {NoStop}%
\bibitem [{\citenamefont {Bizon}\ and\ \citenamefont
  {Chmaj}(1998)}]{Bizon:1998kq}%
  \BibitemOpen
  \bibfield  {author} {\bibinfo {author} {\bibfnamefont {P.}~\bibnamefont
  {Bizon}}\ and\ \bibinfo {author} {\bibfnamefont {T.}~\bibnamefont {Chmaj}},\
  }\bibfield  {title} {\bibinfo {title} {{Critical collapse of Skyrmions}},\
  }\href {https://doi.org/10.1103/PhysRevD.58.041501} {\bibfield  {journal}
  {\bibinfo  {journal} {Phys. Rev. D}\ }\textbf {\bibinfo {volume} {58}},\
  \bibinfo {pages} {041501} (\bibinfo {year} {1998})},\ \Eprint
  {https://arxiv.org/abs/gr-qc/9801012} {arXiv:gr-qc/9801012} \BibitemShut
  {NoStop}%
\bibitem [{\citenamefont {Gundlach}\ and\ \citenamefont
  {Martin-Garcia}(2007)}]{Gundlach:2007gc}%
  \BibitemOpen
  \bibfield  {author} {\bibinfo {author} {\bibfnamefont {C.}~\bibnamefont
  {Gundlach}}\ and\ \bibinfo {author} {\bibfnamefont {J.~M.}\ \bibnamefont
  {Martin-Garcia}},\ }\bibfield  {title} {\bibinfo {title} {{Critical phenomena
  in gravitational collapse}},\ }\href {https://doi.org/10.12942/lrr-2007-5}
  {\bibfield  {journal} {\bibinfo  {journal} {Living Rev. Rel.}\ }\textbf
  {\bibinfo {volume} {10}},\ \bibinfo {pages} {5} (\bibinfo {year} {2007})},\
  \Eprint {https://arxiv.org/abs/0711.4620} {arXiv:0711.4620 [gr-qc]}
  \BibitemShut {NoStop}%
\bibitem [{\citenamefont {Chen}\ \emph
  {et~al.}(2023{\natexlab{b}})\citenamefont {Chen}, \citenamefont {Liu},
  \citenamefont {Tian}, \citenamefont {Wu},\ and\ \citenamefont
  {Zhang}}]{Chen:2022tfy}%
  \BibitemOpen
  \bibfield  {author} {\bibinfo {author} {\bibfnamefont {Q.}~\bibnamefont
  {Chen}}, \bibinfo {author} {\bibfnamefont {Y.}~\bibnamefont {Liu}}, \bibinfo
  {author} {\bibfnamefont {Y.}~\bibnamefont {Tian}}, \bibinfo {author}
  {\bibfnamefont {X.}~\bibnamefont {Wu}},\ and\ \bibinfo {author}
  {\bibfnamefont {H.}~\bibnamefont {Zhang}},\ }\bibfield  {title} {\bibinfo
  {title} {{Quench dynamics in holographic first-order phase transition}},\
  }\href {https://doi.org/10.1103/PhysRevD.108.106017} {\bibfield  {journal}
  {\bibinfo  {journal} {Phys. Rev. D}\ }\textbf {\bibinfo {volume} {108}},\
  \bibinfo {pages} {106017} (\bibinfo {year} {2023}{\natexlab{b}})},\ \Eprint
  {https://arxiv.org/abs/2211.11291} {arXiv:2211.11291 [hep-th]} \BibitemShut
  {NoStop}%
\bibitem [{\citenamefont {Kibble}(1980)}]{Kibble:1980mv}%
  \BibitemOpen
  \bibfield  {author} {\bibinfo {author} {\bibfnamefont {T.~W.~B.}\
  \bibnamefont {Kibble}},\ }\bibfield  {title} {\bibinfo {title} {{Some
  Implications of a Cosmological Phase Transition}},\ }\href
  {https://doi.org/10.1016/0370-1573(80)90091-5} {\bibfield  {journal}
  {\bibinfo  {journal} {Phys. Rept.}\ }\textbf {\bibinfo {volume} {67}},\
  \bibinfo {pages} {183} (\bibinfo {year} {1980})}\BibitemShut {NoStop}%
\bibitem [{\citenamefont {Bigazzi}\ \emph {et~al.}(2020)\citenamefont
  {Bigazzi}, \citenamefont {Caddeo}, \citenamefont {Cotrone},\ and\
  \citenamefont {Paredes}}]{Bigazzi:2020phm}%
  \BibitemOpen
  \bibfield  {author} {\bibinfo {author} {\bibfnamefont {F.}~\bibnamefont
  {Bigazzi}}, \bibinfo {author} {\bibfnamefont {A.}~\bibnamefont {Caddeo}},
  \bibinfo {author} {\bibfnamefont {A.~L.}\ \bibnamefont {Cotrone}},\ and\
  \bibinfo {author} {\bibfnamefont {A.}~\bibnamefont {Paredes}},\ }\bibfield
  {title} {\bibinfo {title} {{Fate of false vacua in holographic first-order
  phase transitions}},\ }\href {https://doi.org/10.1007/JHEP12(2020)200}
  {\bibfield  {journal} {\bibinfo  {journal} {JHEP}\ }\textbf {\bibinfo
  {volume} {12}},\ \bibinfo {pages} {200}},\ \Eprint
  {https://arxiv.org/abs/2008.02579} {arXiv:2008.02579 [hep-th]} \BibitemShut
  {NoStop}%
\bibitem [{\citenamefont {Hindmarsh}\ \emph {et~al.}(2021)\citenamefont
  {Hindmarsh}, \citenamefont {L\"uben}, \citenamefont {Lumma},\ and\
  \citenamefont {Pauly}}]{Hindmarsh:2020hop}%
  \BibitemOpen
  \bibfield  {author} {\bibinfo {author} {\bibfnamefont {M.~B.}\ \bibnamefont
  {Hindmarsh}}, \bibinfo {author} {\bibfnamefont {M.}~\bibnamefont {L\"uben}},
  \bibinfo {author} {\bibfnamefont {J.}~\bibnamefont {Lumma}},\ and\ \bibinfo
  {author} {\bibfnamefont {M.}~\bibnamefont {Pauly}},\ }\bibfield  {title}
  {\bibinfo {title} {{Phase transitions in the early universe}},\ }\href
  {https://doi.org/10.21468/SciPostPhysLectNotes.24} {\bibfield  {journal}
  {\bibinfo  {journal} {SciPost Phys. Lect. Notes}\ }\textbf {\bibinfo {volume}
  {24}},\ \bibinfo {pages} {1} (\bibinfo {year} {2021})},\ \Eprint
  {https://arxiv.org/abs/2008.09136} {arXiv:2008.09136 [astro-ph.CO]}
  \BibitemShut {NoStop}%
\bibitem [{\citenamefont {Bigazzi}\ \emph
  {et~al.}(2021{\natexlab{a}})\citenamefont {Bigazzi}, \citenamefont {Caddeo},
  \citenamefont {Cotrone},\ and\ \citenamefont {Paredes}}]{Bigazzi:2020avc}%
  \BibitemOpen
  \bibfield  {author} {\bibinfo {author} {\bibfnamefont {F.}~\bibnamefont
  {Bigazzi}}, \bibinfo {author} {\bibfnamefont {A.}~\bibnamefont {Caddeo}},
  \bibinfo {author} {\bibfnamefont {A.~L.}\ \bibnamefont {Cotrone}},\ and\
  \bibinfo {author} {\bibfnamefont {A.}~\bibnamefont {Paredes}},\ }\bibfield
  {title} {\bibinfo {title} {{Dark Holograms and Gravitational Waves}},\ }\href
  {https://doi.org/10.1007/JHEP04(2021)094} {\bibfield  {journal} {\bibinfo
  {journal} {JHEP}\ }\textbf {\bibinfo {volume} {04}},\ \bibinfo {pages}
  {094}},\ \Eprint {https://arxiv.org/abs/2011.08757} {arXiv:2011.08757
  [hep-ph]} \BibitemShut {NoStop}%
\bibitem [{\citenamefont {Ares}\ \emph {et~al.}(2020)\citenamefont {Ares},
  \citenamefont {Hindmarsh}, \citenamefont {Hoyos},\ and\ \citenamefont
  {Jokela}}]{Ares:2020lbt}%
  \BibitemOpen
  \bibfield  {author} {\bibinfo {author} {\bibfnamefont {F.~R.}\ \bibnamefont
  {Ares}}, \bibinfo {author} {\bibfnamefont {M.}~\bibnamefont {Hindmarsh}},
  \bibinfo {author} {\bibfnamefont {C.}~\bibnamefont {Hoyos}},\ and\ \bibinfo
  {author} {\bibfnamefont {N.}~\bibnamefont {Jokela}},\ }\bibfield  {title}
  {\bibinfo {title} {{Gravitational waves from a holographic phase
  transition}},\ }\href {https://doi.org/10.1007/JHEP04(2021)100} {\bibfield
  {journal} {\bibinfo  {journal} {JHEP}\ }\textbf {\bibinfo {volume} {21}},\
  \bibinfo {pages} {100}},\ \Eprint {https://arxiv.org/abs/2011.12878}
  {arXiv:2011.12878 [hep-th]} \BibitemShut {NoStop}%
\bibitem [{\citenamefont {Ares}\ \emph {et~al.}(2022)\citenamefont {Ares},
  \citenamefont {Henriksson}, \citenamefont {Hindmarsh}, \citenamefont
  {Hoyos},\ and\ \citenamefont {Jokela}}]{Ares:2021nap}%
  \BibitemOpen
  \bibfield  {author} {\bibinfo {author} {\bibfnamefont {F.~R.}\ \bibnamefont
  {Ares}}, \bibinfo {author} {\bibfnamefont {O.}~\bibnamefont {Henriksson}},
  \bibinfo {author} {\bibfnamefont {M.}~\bibnamefont {Hindmarsh}}, \bibinfo
  {author} {\bibfnamefont {C.}~\bibnamefont {Hoyos}},\ and\ \bibinfo {author}
  {\bibfnamefont {N.}~\bibnamefont {Jokela}},\ }\bibfield  {title} {\bibinfo
  {title} {{Gravitational Waves at Strong Coupling from an Effective Action}},\
  }\href {https://doi.org/10.1103/PhysRevLett.128.131101} {\bibfield  {journal}
  {\bibinfo  {journal} {Phys. Rev. Lett.}\ }\textbf {\bibinfo {volume} {128}},\
  \bibinfo {pages} {131101} (\bibinfo {year} {2022})},\ \Eprint
  {https://arxiv.org/abs/2110.14442} {arXiv:2110.14442 [hep-th]} \BibitemShut
  {NoStop}%
\bibitem [{\citenamefont {Bea}\ \emph {et~al.}(2021{\natexlab{b}})\citenamefont
  {Bea}, \citenamefont {Casalderrey-Solana}, \citenamefont {Giannakopoulos},
  \citenamefont {Mateos}, \citenamefont {Sanchez-Garitaonandia},\ and\
  \citenamefont {Zilh\~ao}}]{Bea:2021zsu}%
  \BibitemOpen
  \bibfield  {author} {\bibinfo {author} {\bibfnamefont {Y.}~\bibnamefont
  {Bea}}, \bibinfo {author} {\bibfnamefont {J.}~\bibnamefont
  {Casalderrey-Solana}}, \bibinfo {author} {\bibfnamefont {T.}~\bibnamefont
  {Giannakopoulos}}, \bibinfo {author} {\bibfnamefont {D.}~\bibnamefont
  {Mateos}}, \bibinfo {author} {\bibfnamefont {M.}~\bibnamefont
  {Sanchez-Garitaonandia}},\ and\ \bibinfo {author} {\bibfnamefont
  {M.}~\bibnamefont {Zilh\~ao}},\ }\bibfield  {title} {\bibinfo {title}
  {{Bubble wall velocity from holography}},\ }\href
  {https://doi.org/10.1103/PhysRevD.104.L121903} {\bibfield  {journal}
  {\bibinfo  {journal} {Phys. Rev. D}\ }\textbf {\bibinfo {volume} {104}},\
  \bibinfo {pages} {L121903} (\bibinfo {year} {2021}{\natexlab{b}})},\ \Eprint
  {https://arxiv.org/abs/2104.05708} {arXiv:2104.05708 [hep-th]} \BibitemShut
  {NoStop}%
\bibitem [{\citenamefont {Bigazzi}\ \emph
  {et~al.}(2021{\natexlab{b}})\citenamefont {Bigazzi}, \citenamefont {Caddeo},
  \citenamefont {Canneti},\ and\ \citenamefont {Cotrone}}]{Bigazzi:2021ucw}%
  \BibitemOpen
  \bibfield  {author} {\bibinfo {author} {\bibfnamefont {F.}~\bibnamefont
  {Bigazzi}}, \bibinfo {author} {\bibfnamefont {A.}~\bibnamefont {Caddeo}},
  \bibinfo {author} {\bibfnamefont {T.}~\bibnamefont {Canneti}},\ and\ \bibinfo
  {author} {\bibfnamefont {A.~L.}\ \bibnamefont {Cotrone}},\ }\bibfield
  {title} {\bibinfo {title} {{Bubble wall velocity at strong coupling}},\
  }\href {https://doi.org/10.1007/JHEP08(2021)090} {\bibfield  {journal}
  {\bibinfo  {journal} {JHEP}\ }\textbf {\bibinfo {volume} {08}},\ \bibinfo
  {pages} {090}},\ \Eprint {https://arxiv.org/abs/2104.12817} {arXiv:2104.12817
  [hep-ph]} \BibitemShut {NoStop}%
\bibitem [{\citenamefont {Janik}\ \emph {et~al.}(2021)\citenamefont {Janik},
  \citenamefont {Jarvinen},\ and\ \citenamefont
  {Sonnenschein}}]{Janik:2021jbq}%
  \BibitemOpen
  \bibfield  {author} {\bibinfo {author} {\bibfnamefont {R.~A.}\ \bibnamefont
  {Janik}}, \bibinfo {author} {\bibfnamefont {M.}~\bibnamefont {Jarvinen}},\
  and\ \bibinfo {author} {\bibfnamefont {J.}~\bibnamefont {Sonnenschein}},\
  }\bibfield  {title} {\bibinfo {title} {{A simple description of holographic
  domain walls in confining theories \textemdash{} extended hydrodynamics}},\
  }\href {https://doi.org/10.1007/JHEP09(2021)129} {\bibfield  {journal}
  {\bibinfo  {journal} {JHEP}\ }\textbf {\bibinfo {volume} {09}},\ \bibinfo
  {pages} {129}},\ \Eprint {https://arxiv.org/abs/2106.02642} {arXiv:2106.02642
  [hep-th]} \BibitemShut {NoStop}%
\bibitem [{\citenamefont {Janik}\ \emph {et~al.}(2022)\citenamefont {Janik},
  \citenamefont {Jarvinen}, \citenamefont {Soltanpanahi},\ and\ \citenamefont
  {Sonnenschein}}]{Janik:2022wsx}%
  \BibitemOpen
  \bibfield  {author} {\bibinfo {author} {\bibfnamefont {R.~A.}\ \bibnamefont
  {Janik}}, \bibinfo {author} {\bibfnamefont {M.}~\bibnamefont {Jarvinen}},
  \bibinfo {author} {\bibfnamefont {H.}~\bibnamefont {Soltanpanahi}},\ and\
  \bibinfo {author} {\bibfnamefont {J.}~\bibnamefont {Sonnenschein}},\
  }\bibfield  {title} {\bibinfo {title} {{Perfect Fluid Hydrodynamic Picture of
  Domain Wall Velocities at Strong Coupling}},\ }\href
  {https://doi.org/10.1103/PhysRevLett.129.081601} {\bibfield  {journal}
  {\bibinfo  {journal} {Phys. Rev. Lett.}\ }\textbf {\bibinfo {volume} {129}},\
  \bibinfo {pages} {081601} (\bibinfo {year} {2022})},\ \Eprint
  {https://arxiv.org/abs/2205.06274} {arXiv:2205.06274 [hep-th]} \BibitemShut
  {NoStop}%
\bibitem [{\citenamefont {Bea}\ \emph {et~al.}(2021{\natexlab{c}})\citenamefont
  {Bea}, \citenamefont {Casalderrey-Solana}, \citenamefont {Giannakopoulos},
  \citenamefont {Jansen}, \citenamefont {Krippendorf}, \citenamefont {Mateos},
  \citenamefont {Sanchez-Garitaonandia},\ and\ \citenamefont
  {Zilh\~ao}}]{Bea:2021zol}%
  \BibitemOpen
  \bibfield  {author} {\bibinfo {author} {\bibfnamefont {Y.}~\bibnamefont
  {Bea}}, \bibinfo {author} {\bibfnamefont {J.}~\bibnamefont
  {Casalderrey-Solana}}, \bibinfo {author} {\bibfnamefont {T.}~\bibnamefont
  {Giannakopoulos}}, \bibinfo {author} {\bibfnamefont {A.}~\bibnamefont
  {Jansen}}, \bibinfo {author} {\bibfnamefont {S.}~\bibnamefont {Krippendorf}},
  \bibinfo {author} {\bibfnamefont {D.}~\bibnamefont {Mateos}}, \bibinfo
  {author} {\bibfnamefont {M.}~\bibnamefont {Sanchez-Garitaonandia}},\ and\
  \bibinfo {author} {\bibfnamefont {M.}~\bibnamefont {Zilh\~ao}},\ }\bibfield
  {title} {\bibinfo {title} {{Spinodal Gravitational Waves}},\ }\href@noop {}
  {\  (\bibinfo {year} {2021}{\natexlab{c}})},\ \Eprint
  {https://arxiv.org/abs/2112.15478} {arXiv:2112.15478 [hep-th]} \BibitemShut
  {NoStop}%
\bibitem [{\citenamefont {Casalderrey-Solana}\ \emph
  {et~al.}(2014{\natexlab{a}})\citenamefont {Casalderrey-Solana}, \citenamefont
  {Liu}, \citenamefont {Mateos}, \citenamefont {Rajagopal},\ and\ \citenamefont
  {Wiedemann}}]{Casalderrey-Solana:2011dxg}%
  \BibitemOpen
  \bibfield  {author} {\bibinfo {author} {\bibfnamefont {J.}~\bibnamefont
  {Casalderrey-Solana}}, \bibinfo {author} {\bibfnamefont {H.}~\bibnamefont
  {Liu}}, \bibinfo {author} {\bibfnamefont {D.}~\bibnamefont {Mateos}},
  \bibinfo {author} {\bibfnamefont {K.}~\bibnamefont {Rajagopal}},\ and\
  \bibinfo {author} {\bibfnamefont {U.~A.}\ \bibnamefont {Wiedemann}},\ }\href
  {https://doi.org/10.1017/9781009403504} {\emph {\bibinfo {title}
  {{Gauge/String Duality, Hot QCD and Heavy Ion Collisions}}}}\ (\bibinfo
  {publisher} {Cambridge University Press},\ \bibinfo {year} {2014})\ \Eprint
  {https://arxiv.org/abs/1101.0618} {arXiv:1101.0618 [hep-th]} \BibitemShut
  {NoStop}%
\bibitem [{\citenamefont {Chesler}\ and\ \citenamefont
  {Yaffe}(2011)}]{Chesler:2010bi}%
  \BibitemOpen
  \bibfield  {author} {\bibinfo {author} {\bibfnamefont {P.~M.}\ \bibnamefont
  {Chesler}}\ and\ \bibinfo {author} {\bibfnamefont {L.~G.}\ \bibnamefont
  {Yaffe}},\ }\bibfield  {title} {\bibinfo {title} {{Holography and colliding
  gravitational shock waves in asymptotically AdS$_{5}$ spacetime}},\ }\href
  {https://doi.org/10.1103/PhysRevLett.106.021601} {\bibfield  {journal}
  {\bibinfo  {journal} {Phys. Rev. Lett.}\ }\textbf {\bibinfo {volume} {106}},\
  \bibinfo {pages} {021601} (\bibinfo {year} {2011})},\ \Eprint
  {https://arxiv.org/abs/1011.3562} {arXiv:1011.3562 [hep-th]} \BibitemShut
  {NoStop}%
\bibitem [{\citenamefont {Casalderrey-Solana}\ \emph
  {et~al.}(2013)\citenamefont {Casalderrey-Solana}, \citenamefont {Heller},
  \citenamefont {Mateos},\ and\ \citenamefont {van~der
  Schee}}]{Casalderrey-Solana:2013aba}%
  \BibitemOpen
  \bibfield  {author} {\bibinfo {author} {\bibfnamefont {J.}~\bibnamefont
  {Casalderrey-Solana}}, \bibinfo {author} {\bibfnamefont {M.~P.}\ \bibnamefont
  {Heller}}, \bibinfo {author} {\bibfnamefont {D.}~\bibnamefont {Mateos}},\
  and\ \bibinfo {author} {\bibfnamefont {W.}~\bibnamefont {van~der Schee}},\
  }\bibfield  {title} {\bibinfo {title} {{From full stopping to transparency in
  a holographic model of heavy ion collisions}},\ }\href
  {https://doi.org/10.1103/PhysRevLett.111.181601} {\bibfield  {journal}
  {\bibinfo  {journal} {Phys. Rev. Lett.}\ }\textbf {\bibinfo {volume} {111}},\
  \bibinfo {pages} {181601} (\bibinfo {year} {2013})},\ \Eprint
  {https://arxiv.org/abs/1305.4919} {arXiv:1305.4919 [hep-th]} \BibitemShut
  {NoStop}%
\bibitem [{\citenamefont {Casalderrey-Solana}\ \emph
  {et~al.}(2014{\natexlab{b}})\citenamefont {Casalderrey-Solana}, \citenamefont
  {Heller}, \citenamefont {Mateos},\ and\ \citenamefont {van~der
  Schee}}]{Casalderrey-Solana:2013sxa}%
  \BibitemOpen
  \bibfield  {author} {\bibinfo {author} {\bibfnamefont {J.}~\bibnamefont
  {Casalderrey-Solana}}, \bibinfo {author} {\bibfnamefont {M.~P.}\ \bibnamefont
  {Heller}}, \bibinfo {author} {\bibfnamefont {D.}~\bibnamefont {Mateos}},\
  and\ \bibinfo {author} {\bibfnamefont {W.}~\bibnamefont {van~der Schee}},\
  }\bibfield  {title} {\bibinfo {title} {{Longitudinal Coherence in a
  Holographic Model of Asymmetric Collisions}},\ }\href
  {https://doi.org/10.1103/PhysRevLett.112.221602} {\bibfield  {journal}
  {\bibinfo  {journal} {Phys. Rev. Lett.}\ }\textbf {\bibinfo {volume} {112}},\
  \bibinfo {pages} {221602} (\bibinfo {year} {2014}{\natexlab{b}})},\ \Eprint
  {https://arxiv.org/abs/1312.2956} {arXiv:1312.2956 [hep-th]} \BibitemShut
  {NoStop}%
\bibitem [{\citenamefont {Chesler}\ and\ \citenamefont
  {Yaffe}(2015)}]{Chesler:2015wra}%
  \BibitemOpen
  \bibfield  {author} {\bibinfo {author} {\bibfnamefont {P.~M.}\ \bibnamefont
  {Chesler}}\ and\ \bibinfo {author} {\bibfnamefont {L.~G.}\ \bibnamefont
  {Yaffe}},\ }\bibfield  {title} {\bibinfo {title} {{Holography and off-center
  collisions of localized shock waves}},\ }\href
  {https://doi.org/10.1007/JHEP10(2015)070} {\bibfield  {journal} {\bibinfo
  {journal} {JHEP}\ }\textbf {\bibinfo {volume} {10}},\ \bibinfo {pages}
  {070}},\ \Eprint {https://arxiv.org/abs/1501.04644} {arXiv:1501.04644
  [hep-th]} \BibitemShut {NoStop}%
\bibitem [{\citenamefont {Chesler}(2015)}]{Chesler:2015bba}%
  \BibitemOpen
  \bibfield  {author} {\bibinfo {author} {\bibfnamefont {P.~M.}\ \bibnamefont
  {Chesler}},\ }\bibfield  {title} {\bibinfo {title} {{Colliding shock waves
  and hydrodynamics in small systems}},\ }\href
  {https://doi.org/10.1103/PhysRevLett.115.241602} {\bibfield  {journal}
  {\bibinfo  {journal} {Phys. Rev. Lett.}\ }\textbf {\bibinfo {volume} {115}},\
  \bibinfo {pages} {241602} (\bibinfo {year} {2015})},\ \Eprint
  {https://arxiv.org/abs/1506.02209} {arXiv:1506.02209 [hep-th]} \BibitemShut
  {NoStop}%
\bibitem [{\citenamefont {Chesler}(2016)}]{Chesler:2016ceu}%
  \BibitemOpen
  \bibfield  {author} {\bibinfo {author} {\bibfnamefont {P.~M.}\ \bibnamefont
  {Chesler}},\ }\bibfield  {title} {\bibinfo {title} {{How big are the smallest
  drops of quark-gluon plasma?}},\ }\href
  {https://doi.org/10.1007/JHEP03(2016)146} {\bibfield  {journal} {\bibinfo
  {journal} {JHEP}\ }\textbf {\bibinfo {volume} {03}},\ \bibinfo {pages}
  {146}},\ \Eprint {https://arxiv.org/abs/1601.01583} {arXiv:1601.01583
  [hep-th]} \BibitemShut {NoStop}%
\bibitem [{\citenamefont {Attems}\ \emph
  {et~al.}(2017{\natexlab{b}})\citenamefont {Attems}, \citenamefont
  {Casalderrey-Solana}, \citenamefont {Mateos}, \citenamefont
  {Santos-Oliv\'an}, \citenamefont {Sopuerta}, \citenamefont {Triana},\ and\
  \citenamefont {Zilh\~ao}}]{Attems:2016tby}%
  \BibitemOpen
  \bibfield  {author} {\bibinfo {author} {\bibfnamefont {M.}~\bibnamefont
  {Attems}}, \bibinfo {author} {\bibfnamefont {J.}~\bibnamefont
  {Casalderrey-Solana}}, \bibinfo {author} {\bibfnamefont {D.}~\bibnamefont
  {Mateos}}, \bibinfo {author} {\bibfnamefont {D.}~\bibnamefont
  {Santos-Oliv\'an}}, \bibinfo {author} {\bibfnamefont {C.~F.}\ \bibnamefont
  {Sopuerta}}, \bibinfo {author} {\bibfnamefont {M.}~\bibnamefont {Triana}},\
  and\ \bibinfo {author} {\bibfnamefont {M.}~\bibnamefont {Zilh\~ao}},\
  }\bibfield  {title} {\bibinfo {title} {{Holographic Collisions in
  Non-conformal Theories}},\ }\href {https://doi.org/10.1007/JHEP01(2017)026}
  {\bibfield  {journal} {\bibinfo  {journal} {JHEP}\ }\textbf {\bibinfo
  {volume} {01}},\ \bibinfo {pages} {026}},\ \Eprint
  {https://arxiv.org/abs/1604.06439} {arXiv:1604.06439 [hep-th]} \BibitemShut
  {NoStop}%
\bibitem [{\citenamefont {Attems}\ \emph {et~al.}(2018)\citenamefont {Attems},
  \citenamefont {Bea}, \citenamefont {Casalderrey-Solana}, \citenamefont
  {Mateos}, \citenamefont {Triana},\ and\ \citenamefont
  {Zilh\~ao}}]{Attems:2018gou}%
  \BibitemOpen
  \bibfield  {author} {\bibinfo {author} {\bibfnamefont {M.}~\bibnamefont
  {Attems}}, \bibinfo {author} {\bibfnamefont {Y.}~\bibnamefont {Bea}},
  \bibinfo {author} {\bibfnamefont {J.}~\bibnamefont {Casalderrey-Solana}},
  \bibinfo {author} {\bibfnamefont {D.}~\bibnamefont {Mateos}}, \bibinfo
  {author} {\bibfnamefont {M.}~\bibnamefont {Triana}},\ and\ \bibinfo {author}
  {\bibfnamefont {M.}~\bibnamefont {Zilh\~ao}},\ }\bibfield  {title} {\bibinfo
  {title} {{Holographic Collisions across a Phase Transition}},\ }\href
  {https://doi.org/10.1103/PhysRevLett.121.261601} {\bibfield  {journal}
  {\bibinfo  {journal} {Phys. Rev. Lett.}\ }\textbf {\bibinfo {volume} {121}},\
  \bibinfo {pages} {261601} (\bibinfo {year} {2018})},\ \Eprint
  {https://arxiv.org/abs/1807.05175} {arXiv:1807.05175 [hep-th]} \BibitemShut
  {NoStop}%
\bibitem [{\citenamefont {Bea}\ \emph {et~al.}(2022{\natexlab{b}})\citenamefont
  {Bea}, \citenamefont {Casalderrey-Solana}, \citenamefont {Giannakopoulos},
  \citenamefont {Mateos}, \citenamefont {Sanchez-Garitaonandia},\ and\
  \citenamefont {Zilh\~ao}}]{Bea:2021ieq}%
  \BibitemOpen
  \bibfield  {author} {\bibinfo {author} {\bibfnamefont {Y.}~\bibnamefont
  {Bea}}, \bibinfo {author} {\bibfnamefont {J.}~\bibnamefont
  {Casalderrey-Solana}}, \bibinfo {author} {\bibfnamefont {T.}~\bibnamefont
  {Giannakopoulos}}, \bibinfo {author} {\bibfnamefont {D.}~\bibnamefont
  {Mateos}}, \bibinfo {author} {\bibfnamefont {M.}~\bibnamefont
  {Sanchez-Garitaonandia}},\ and\ \bibinfo {author} {\bibfnamefont
  {M.}~\bibnamefont {Zilh\~ao}},\ }\bibfield  {title} {\bibinfo {title}
  {{Domain collisions}},\ }\href {https://doi.org/10.1007/JHEP06(2022)025}
  {\bibfield  {journal} {\bibinfo  {journal} {JHEP}\ }\textbf {\bibinfo
  {volume} {06}},\ \bibinfo {pages} {025}},\ \Eprint
  {https://arxiv.org/abs/2111.03355} {arXiv:2111.03355 [hep-th]} \BibitemShut
  {NoStop}%
\bibitem [{\citenamefont {Adams}\ \emph {et~al.}(2013)\citenamefont {Adams},
  \citenamefont {Chesler},\ and\ \citenamefont {Liu}}]{Adams:2012pj}%
  \BibitemOpen
  \bibfield  {author} {\bibinfo {author} {\bibfnamefont {A.}~\bibnamefont
  {Adams}}, \bibinfo {author} {\bibfnamefont {P.~M.}\ \bibnamefont {Chesler}},\
  and\ \bibinfo {author} {\bibfnamefont {H.}~\bibnamefont {Liu}},\ }\bibfield
  {title} {\bibinfo {title} {{Holographic Vortex Liquids and Superfluid
  Turbulence}},\ }\href {https://doi.org/10.1126/science.1233529} {\bibfield
  {journal} {\bibinfo  {journal} {Science}\ }\textbf {\bibinfo {volume}
  {341}},\ \bibinfo {pages} {368} (\bibinfo {year} {2013})},\ \Eprint
  {https://arxiv.org/abs/1212.0281} {arXiv:1212.0281 [hep-th]} \BibitemShut
  {NoStop}%
\bibitem [{\citenamefont {Adams}\ \emph {et~al.}(2014)\citenamefont {Adams},
  \citenamefont {Chesler},\ and\ \citenamefont {Liu}}]{Adams:2013vsa}%
  \BibitemOpen
  \bibfield  {author} {\bibinfo {author} {\bibfnamefont {A.}~\bibnamefont
  {Adams}}, \bibinfo {author} {\bibfnamefont {P.~M.}\ \bibnamefont {Chesler}},\
  and\ \bibinfo {author} {\bibfnamefont {H.}~\bibnamefont {Liu}},\ }\bibfield
  {title} {\bibinfo {title} {{Holographic turbulence}},\ }\href
  {https://doi.org/10.1103/PhysRevLett.112.151602} {\bibfield  {journal}
  {\bibinfo  {journal} {Phys. Rev. Lett.}\ }\textbf {\bibinfo {volume} {112}},\
  \bibinfo {pages} {151602} (\bibinfo {year} {2014})},\ \Eprint
  {https://arxiv.org/abs/1307.7267} {arXiv:1307.7267 [hep-th]} \BibitemShut
  {NoStop}%
\bibitem [{\citenamefont {Herzog}\ \emph {et~al.}(2009)\citenamefont {Herzog},
  \citenamefont {Kovtun},\ and\ \citenamefont {Son}}]{Herzog:2008he}%
  \BibitemOpen
  \bibfield  {author} {\bibinfo {author} {\bibfnamefont {C.~P.}\ \bibnamefont
  {Herzog}}, \bibinfo {author} {\bibfnamefont {P.~K.}\ \bibnamefont {Kovtun}},\
  and\ \bibinfo {author} {\bibfnamefont {D.~T.}\ \bibnamefont {Son}},\
  }\bibfield  {title} {\bibinfo {title} {{Holographic model of
  superfluidity}},\ }\href {https://doi.org/10.1103/PhysRevD.79.066002}
  {\bibfield  {journal} {\bibinfo  {journal} {Phys. Rev. D}\ }\textbf {\bibinfo
  {volume} {79}},\ \bibinfo {pages} {066002} (\bibinfo {year} {2009})},\
  \Eprint {https://arxiv.org/abs/0809.4870} {arXiv:0809.4870 [hep-th]}
  \BibitemShut {NoStop}%
\bibitem [{\citenamefont {Bhaseen}\ \emph {et~al.}(2013)\citenamefont
  {Bhaseen}, \citenamefont {Gauntlett}, \citenamefont {Simons}, \citenamefont
  {Sonner},\ and\ \citenamefont {Wiseman}}]{Bhaseen:2012gg}%
  \BibitemOpen
  \bibfield  {author} {\bibinfo {author} {\bibfnamefont {M.~J.}\ \bibnamefont
  {Bhaseen}}, \bibinfo {author} {\bibfnamefont {J.~P.}\ \bibnamefont
  {Gauntlett}}, \bibinfo {author} {\bibfnamefont {B.~D.}\ \bibnamefont
  {Simons}}, \bibinfo {author} {\bibfnamefont {J.}~\bibnamefont {Sonner}},\
  and\ \bibinfo {author} {\bibfnamefont {T.}~\bibnamefont {Wiseman}},\
  }\bibfield  {title} {\bibinfo {title} {{Holographic Superfluids and the
  Dynamics of Symmetry Breaking}},\ }\href
  {https://doi.org/10.1103/PhysRevLett.110.015301} {\bibfield  {journal}
  {\bibinfo  {journal} {Phys. Rev. Lett.}\ }\textbf {\bibinfo {volume} {110}},\
  \bibinfo {pages} {015301} (\bibinfo {year} {2013})},\ \Eprint
  {https://arxiv.org/abs/1207.4194} {arXiv:1207.4194 [hep-th]} \BibitemShut
  {NoStop}%
\bibitem [{\citenamefont {Zhao}\ \emph {et~al.}(2024)\citenamefont {Zhao},
  \citenamefont {Nie}, \citenamefont {Zhao}, \citenamefont {Zeng},
  \citenamefont {Tian},\ and\ \citenamefont {Baggioli}}]{Zhao:2023ffs}%
  \BibitemOpen
  \bibfield  {author} {\bibinfo {author} {\bibfnamefont {X.}~\bibnamefont
  {Zhao}}, \bibinfo {author} {\bibfnamefont {Z.-Y.}\ \bibnamefont {Nie}},
  \bibinfo {author} {\bibfnamefont {Z.-Q.}\ \bibnamefont {Zhao}}, \bibinfo
  {author} {\bibfnamefont {H.-B.}\ \bibnamefont {Zeng}}, \bibinfo {author}
  {\bibfnamefont {Y.}~\bibnamefont {Tian}},\ and\ \bibinfo {author}
  {\bibfnamefont {M.}~\bibnamefont {Baggioli}},\ }\bibfield  {title} {\bibinfo
  {title} {{Dynamical evolution of spinodal decomposition in holographic
  superfluids}},\ }\href {https://doi.org/10.1007/JHEP02(2024)184} {\bibfield
  {journal} {\bibinfo  {journal} {JHEP}\ }\textbf {\bibinfo {volume} {02}},\
  \bibinfo {pages} {184}},\ \Eprint {https://arxiv.org/abs/2311.08277}
  {arXiv:2311.08277 [hep-th]} \BibitemShut {NoStop}%
\bibitem [{\citenamefont {Bhattacharyya}\ \emph {et~al.}(2008)\citenamefont
  {Bhattacharyya}, \citenamefont {Hubeny}, \citenamefont {Minwalla},\ and\
  \citenamefont {Rangamani}}]{Bhattacharyya:2007vjd}%
  \BibitemOpen
  \bibfield  {author} {\bibinfo {author} {\bibfnamefont {S.}~\bibnamefont
  {Bhattacharyya}}, \bibinfo {author} {\bibfnamefont {V.~E.}\ \bibnamefont
  {Hubeny}}, \bibinfo {author} {\bibfnamefont {S.}~\bibnamefont {Minwalla}},\
  and\ \bibinfo {author} {\bibfnamefont {M.}~\bibnamefont {Rangamani}},\
  }\bibfield  {title} {\bibinfo {title} {{Nonlinear Fluid Dynamics from
  Gravity}},\ }\href {https://doi.org/10.1088/1126-6708/2008/02/045} {\bibfield
   {journal} {\bibinfo  {journal} {JHEP}\ }\textbf {\bibinfo {volume} {02}},\
  \bibinfo {pages} {045}},\ \Eprint {https://arxiv.org/abs/0712.2456}
  {arXiv:0712.2456 [hep-th]} \BibitemShut {NoStop}%
\bibitem [{\citenamefont {Ecker}\ \emph {et~al.}(2021)\citenamefont {Ecker},
  \citenamefont {Erdmenger},\ and\ \citenamefont {van~der
  Schee}}]{Ecker:2021ukv}%
  \BibitemOpen
  \bibfield  {author} {\bibinfo {author} {\bibfnamefont {C.}~\bibnamefont
  {Ecker}}, \bibinfo {author} {\bibfnamefont {J.}~\bibnamefont {Erdmenger}},\
  and\ \bibinfo {author} {\bibfnamefont {W.}~\bibnamefont {van~der Schee}},\
  }\bibfield  {title} {\bibinfo {title} {{Non-equilibrium steady state
  formation in 3+1 dimensions}},\ }\href
  {https://doi.org/10.21468/SciPostPhys.11.3.047} {\bibfield  {journal}
  {\bibinfo  {journal} {SciPost Phys.}\ }\textbf {\bibinfo {volume} {11}},\
  \bibinfo {pages} {047} (\bibinfo {year} {2021})},\ \Eprint
  {https://arxiv.org/abs/2103.10435} {arXiv:2103.10435 [hep-th]} \BibitemShut
  {NoStop}%
\bibitem [{\citenamefont {Gibbons}\ and\ \citenamefont
  {Hawking}(1977)}]{Gibbons:1976ue}%
  \BibitemOpen
  \bibfield  {author} {\bibinfo {author} {\bibfnamefont {G.~W.}\ \bibnamefont
  {Gibbons}}\ and\ \bibinfo {author} {\bibfnamefont {S.~W.}\ \bibnamefont
  {Hawking}},\ }\bibfield  {title} {\bibinfo {title} {{Action Integrals and
  Partition Functions in Quantum Gravity}},\ }\href
  {https://doi.org/10.1103/PhysRevD.15.2752} {\bibfield  {journal} {\bibinfo
  {journal} {Phys. Rev. D}\ }\textbf {\bibinfo {volume} {15}},\ \bibinfo
  {pages} {2752} (\bibinfo {year} {1977})}\BibitemShut {NoStop}%
\bibitem [{\citenamefont {Bianchi}\ \emph {et~al.}(2002)\citenamefont
  {Bianchi}, \citenamefont {Freedman},\ and\ \citenamefont
  {Skenderis}}]{Bianchi:2001kw}%
  \BibitemOpen
  \bibfield  {author} {\bibinfo {author} {\bibfnamefont {M.}~\bibnamefont
  {Bianchi}}, \bibinfo {author} {\bibfnamefont {D.~Z.}\ \bibnamefont
  {Freedman}},\ and\ \bibinfo {author} {\bibfnamefont {K.}~\bibnamefont
  {Skenderis}},\ }\bibfield  {title} {\bibinfo {title} {{Holographic
  renormalization}},\ }\href {https://doi.org/10.1016/S0550-3213(02)00179-7}
  {\bibfield  {journal} {\bibinfo  {journal} {Nucl. Phys. B}\ }\textbf
  {\bibinfo {volume} {631}},\ \bibinfo {pages} {159} (\bibinfo {year}
  {2002})},\ \Eprint {https://arxiv.org/abs/hep-th/0112119}
  {arXiv:hep-th/0112119} \BibitemShut {NoStop}%
\bibitem [{\citenamefont {Elvang}\ and\ \citenamefont
  {Hadjiantonis}(2016)}]{Elvang:2016tzz}%
  \BibitemOpen
  \bibfield  {author} {\bibinfo {author} {\bibfnamefont {H.}~\bibnamefont
  {Elvang}}\ and\ \bibinfo {author} {\bibfnamefont {M.}~\bibnamefont
  {Hadjiantonis}},\ }\bibfield  {title} {\bibinfo {title} {{A Practical
  Approach to the Hamilton-Jacobi Formulation of Holographic
  Renormalization}},\ }\href {https://doi.org/10.1007/JHEP06(2016)046}
  {\bibfield  {journal} {\bibinfo  {journal} {JHEP}\ }\textbf {\bibinfo
  {volume} {06}},\ \bibinfo {pages} {046}},\ \Eprint
  {https://arxiv.org/abs/1603.04485} {arXiv:1603.04485 [hep-th]} \BibitemShut
  {NoStop}%
\bibitem [{\citenamefont {Cai}\ \emph {et~al.}(2022)\citenamefont {Cai},
  \citenamefont {He}, \citenamefont {Li},\ and\ \citenamefont
  {Wang}}]{Cai:2022omk}%
  \BibitemOpen
  \bibfield  {author} {\bibinfo {author} {\bibfnamefont {R.-G.}\ \bibnamefont
  {Cai}}, \bibinfo {author} {\bibfnamefont {S.}~\bibnamefont {He}}, \bibinfo
  {author} {\bibfnamefont {L.}~\bibnamefont {Li}},\ and\ \bibinfo {author}
  {\bibfnamefont {Y.-X.}\ \bibnamefont {Wang}},\ }\bibfield  {title} {\bibinfo
  {title} {{Probing QCD critical point and induced gravitational wave by black
  hole physics}},\ }\href {https://doi.org/10.1103/PhysRevD.106.L121902}
  {\bibfield  {journal} {\bibinfo  {journal} {Phys. Rev. D}\ }\textbf {\bibinfo
  {volume} {106}},\ \bibinfo {pages} {L121902} (\bibinfo {year} {2022})},\
  \Eprint {https://arxiv.org/abs/2201.02004} {arXiv:2201.02004 [hep-th]}
  \BibitemShut {NoStop}%
\bibitem [{\citenamefont {Gursoy}\ \emph {et~al.}(2009)\citenamefont {Gursoy},
  \citenamefont {Kiritsis}, \citenamefont {Mazzanti},\ and\ \citenamefont
  {Nitti}}]{Gursoy:2008za}%
  \BibitemOpen
  \bibfield  {author} {\bibinfo {author} {\bibfnamefont {U.}~\bibnamefont
  {Gursoy}}, \bibinfo {author} {\bibfnamefont {E.}~\bibnamefont {Kiritsis}},
  \bibinfo {author} {\bibfnamefont {L.}~\bibnamefont {Mazzanti}},\ and\
  \bibinfo {author} {\bibfnamefont {F.}~\bibnamefont {Nitti}},\ }\bibfield
  {title} {\bibinfo {title} {{Holography and Thermodynamics of 5D
  Dilaton-gravity}},\ }\href {https://doi.org/10.1088/1126-6708/2009/05/033}
  {\bibfield  {journal} {\bibinfo  {journal} {JHEP}\ }\textbf {\bibinfo
  {volume} {05}},\ \bibinfo {pages} {033}},\ \Eprint
  {https://arxiv.org/abs/0812.0792} {arXiv:0812.0792 [hep-th]} \BibitemShut
  {NoStop}%
\bibitem [{\citenamefont {Li}\ \emph {et~al.}(2023)\citenamefont {Li},
  \citenamefont {Wang},\ and\ \citenamefont {Yuwen}}]{Li:2023xto}%
  \BibitemOpen
  \bibfield  {author} {\bibinfo {author} {\bibfnamefont {L.}~\bibnamefont
  {Li}}, \bibinfo {author} {\bibfnamefont {S.-J.}\ \bibnamefont {Wang}},\ and\
  \bibinfo {author} {\bibfnamefont {Z.-Y.}\ \bibnamefont {Yuwen}},\ }\bibfield
  {title} {\bibinfo {title} {{Bubble expansion at strong coupling}},\ }\href
  {https://doi.org/10.1103/PhysRevD.108.096033} {\bibfield  {journal} {\bibinfo
   {journal} {Phys. Rev. D}\ }\textbf {\bibinfo {volume} {108}},\ \bibinfo
  {pages} {096033} (\bibinfo {year} {2023})},\ \Eprint
  {https://arxiv.org/abs/2302.10042} {arXiv:2302.10042 [hep-th]} \BibitemShut
  {NoStop}%
\end{thebibliography}%

\end{document}